\documentclass[10pt,journal,compsoc]{IEEEtran}
\usepackage{epsfig,endnotes}
\usepackage{multicol,graphicx,amssymb,amsmath,latexsym,color,fancyhdr,url, subfig,multirow,colortbl,listings,color,xspace,hyperref}

\usepackage[normalem]{ulem}
\usepackage[linesnumbered,boxed]{algorithm2e}

\SetKwComment{tcp}{$\triangleright$ }{}

\SetCommentSty{mycommfont}

\newcommand{\framework}{DeepWear\xspace}

\newcommand{\mysection}[1]{\vspace{-.06in}\section{#1}\vspace{-.01in}}
\newcommand{\mysubsection}[1]{\vspace{-.06in}\subsection{#1}\vspace{-.01in}}

\definecolor{dkgreen}{rgb}{0,0.6,0}
\definecolor{gray}{rgb}{0.5,0.5,0.5}
\definecolor{mauve}{rgb}{0.58,0,0.82}

\newcommand{\revise}[1]{{#1}}
\newcommand{\mrev}[1]{{#1}}

\usepackage{array}
\newcolumntype{L}[1]{>{\raggedright\let\newline\\\arraybackslash\hspace{0pt}}m{#1}}
\newcolumntype{C}[1]{>{\centering\let\newline\\\arraybackslash\hspace{0pt}}m{#1}}
\newcolumntype{R}[1]{>{\raggedleft\let\newline\\\arraybackslash\hspace{0pt}}m{#1}}
\newcommand{\argmin}[1]{\underset{#1}{\operatorname{arg}\,\operatorname{min}}\;}

\begin{document}


\title{\framework: Adaptive Local Offloading for On-Wearable Deep Learning}

\author{Mengwei Xu, Feng Qian, Mengze Zhu, Feifan Huang, Saumay Pushp, Xuanzhe Liu~\IEEEmembership{Member, IEEE}

\IEEEcompsocitemizethanks{\IEEEcompsocthanksitem
Mengwei~Xu, Mengze~Zhu, Feifan~Huang, and Xuanzhe~Liu are with the Key
Laboratory of High Confidence Software Technologies (Peking
University), Ministry of Education, Beijing, China, 100871. Email: \{xumengwei, zhumengze, huangfeifan, liuxuanzhe\}@pku.edu.cn }%
\IEEEcompsocitemizethanks{\IEEEcompsocthanksitem
Feng Qian is with the Computer Science and Engineering Department at University of Minnesota -- Twin Cities, 200 Union Street SE, Minneapolis MN 55455. Email: fengqian@umn.edu}
\IEEEcompsocitemizethanks{\IEEEcompsocthanksitem
Saumay Pushp is with Korea Advanced Institute of Science and Technology, 291 Daehak-ro, Eoeun-dong, Yuseong-gu, Daejeon. Email: saumay@nclab.kaist.ac.kr}
}

\markboth{IEEE Transactions on Mobile Computing,~Vol.~XX, No.~XX, XXXX~201X}%
{Xu \MakeLowercase{\textit{et al.}}: \framework: \revise{Enabling} Deep Learning on Wearable Devices via Adaptive Local Offloading }

\IEEEtitleabstractindextext{%
\begin{abstract}
Due to their on-body and ubiquitous nature, wearables can generate a wide range of unique sensor data creating countless opportunities for deep learning tasks.
We propose \framework, a deep learning (DL) framework for wearable devices to improve the performance and reduce the energy footprint.
\framework~strategically offloads DL tasks from a wearable device to its paired handheld device through local network connectivity such as Bluetooth.
Compared to the remote-cloud-based offloading, \framework requires no Internet connectivity, consumes less energy, and is robust to privacy breach.
%
\framework~provides various novel techniques such as
context-aware offloading, strategic model partition, and pipelining support to efficiently utilize the processing capacity from nearby paired handhelds.
%
%
Deployed as a user-space library, \framework offers developer-friendly APIs that are as simple as those in traditional DL libraries such as TensorFlow.
We have implemented \framework on the Android OS and evaluated it on COTS smartphones and smartwatches with real DL models.
\framework brings up to \textbf{5.08X} and \textbf{23.0X} execution speedup, as well as \textbf{53.5\%} and \textbf{85.5\%} energy saving compared to wearable-only and handheld-only strategies, respectively. 
\end{abstract}

\begin{IEEEkeywords}
Wearables; Deep Learning; Offloading
\end{IEEEkeywords}}

\maketitle
\IEEEdisplaynontitleabstractindextext

%
\IEEEpeerreviewmaketitle

\mysection{Introduction}\label{sec:intro}

Making deep learning (DL for short in the rest of this paper) tasks run on mobile devices has raised huge interests in both the academia~\cite{conf/mobisys/MathurLBBFK17, conf/huc/RaduLBMMK16, conf/ipsn/LaneBGFJQK16, conf/mobicom/GeorgievLRM16, conf/sensys/BhattacharyaL16, conf/huc/LaneGQ15, conf/wmcsa/LaneG15, xu2018deeptype} and the industry~\cite{onDeviceIntelligence, GoogleTranslate, xu2018mobile}.
In this paper, we focus on how to effectively and efficiently apply DL on wearable devices. Our study is motivated by three key observations.
\revise{First, wearable devices are becoming increasingly popular. According to a recent market research report, the estimated global market value of smartwatch is \$10.2 billion in 2017, and is expected to witness an annual growth rate of 22.3\% from 2018 to 2023~\cite{smartwatch2023}.}
%
Second, DL on wearable devices enables new applications.
Due to their on-body and ubiquitous nature, wearables can collect a wide spectrum of data such as body gesture, heartbeat reading, fitness tracking, eye tracking, and vision (through a smart glass). Such unique data creates countless applications for DL.
Third, despite a plethora of work on DL on smartphones, so far very few studies focus specifically on the interplay between DL and the wearable ecosystem.

In practice, supporting DL on wearable devices is quite challenging, due to the heavy computation requirements of DL and constrained processing capacity on today's COTS (commercial off-the-shelf) wearable devices.
%
%
Intuitively, running DL tasks locally is not a good option for most wearables. Then an instinct idea
is to perform \emph{offloading}~\cite{MAUI, gordon2015accelerating}.
Instead of offloading computations to the remote cloud, we instantiate the idea of Edge Computing~\cite{ha2014towards} by \emph{offloading DL tasks to a nearby mobile device} (e.g., typically a smartphone or a tablet) that has local connectivity with the wearable. Such a ``local'' offloading is indeed feasible for three reasons. (1) As to be demonstrated in our study, today's handheld devices such as smartphones are sufficiently powerful with multi-core CPU, fast GPU, and GBs of memory.
(2) The vast majority of wearables (e.g., smartwatches and smart glasses) are by default paired with a handheld device and using it as a ``gateway'' to access the external world. For example, a recent user study~\cite{conf/mobisys/LiuCQGLWC17} reports that a smartwatch is paired with a smartphone during 84\% of the day time.
\revise{
(3) Prior efforts have been invested in reducing the computation overhead of DL tasks through various optimizations such as model compression~\cite{conf/ipsn/LaneBGFJQK16,conf/cvpr/WuLWHC16,conf/nips/DentonZBLF14,conf/mobisys/HanSPAWK16}.
In our work, we strategically integrate and instantiate some of their concepts into our practical system to make DL tasks wearable-friendly.
}

%

We envision that such a local (edge) offloading approach has three key advantages.
First, offloading to a handheld does not require the not-always-reliable Internet connectivity that can lead to high energy and monetary cost (e.g., over cellular networks). Instead, the communication between the wearable and the handheld can be realized by cheap short-range radio such as Bluetooth or Bluetooth Low Energy (BLE).
Second, users routinely carry \emph{both} wearables and the paired handheld devices, making offloading ubiquitously feasible.
Third, offloading to paired handhelds minimizes risks of privacy leak because the potentially sensitive data (e.g., medical sensor data) generated from wearables is never leaked to the network.

Motivated by the preceding analysis, we design, implement, and evaluate \framework, a holistic DL framework that supports local offloading for wearable DL applications.
%
\framework~has several salient features as described below.

\noindent $\bullet$ \textbf{Context-aware offloading scheduling}. We make a first in-depth measurement study to demystify the performance of wearable-side DL tasks and reveal the potential improvements that can be gained through offloading. Making an appropriate offloading decision involves scrutinizing a wide range of factors
including the DL model structure, the application's latency requirement, and the network connectivity condition, etc. In addition, our offloading target (the handheld) introduces additional complexities: despite being more powerful than a wearable, a handheld still has limited processing power (compared to the cloud) and battery life; as a personal computing device, a handheld also runs other apps that consume system resources by
incurring bursty workload.
Therefore, \framework further takes into account the status of handheld. We incorporate the preceding considerations into a \textit{lightweight online scheduling algorithm} that judiciously determines which, when, and how to offload.

\noindent $\bullet$ \textbf{Partial offloading}. Instead of making a binary decision of offloading the whole DL model versus executing the entire model locally, \framework supports the \emph{partial offloading}. Specifically, \framework splits a model into two sub-models that are separately executed first on the wearable and then on the handheld. We found that in some scenarios, partial offloading outperforms the binary decision, because an internal layer inside the model may yield a smaller intermediate output compared to the original input size, thus reducing the data transfer delay.
To support the partial offloading, we develop a
heuristic-based algorithm that efficiently identifies a set of candidate partition points whose exhaustive search takes exponential time.
The optimal splitting point can then be quickly determined by examining the small candidate set. Our partial offloading approach can work with any DL model with arbitrary topology.



\noindent $\bullet$ \textbf{Optimized data streaming}. \framework~introduces the additional optimization for streaming input such as video frames and audio snippets continuously fed into the same model.
Specifically, \framework employs \emph{pipelined processing} on wearable and handheld,
which helps fully utilize the computation resources on both devices and thus effectively improves the overall throughput.


\noindent $\bullet$ \textbf{Application transparency and good usability}. We propose a modular design of the \framework~system whose most logic is transparent to the application. Developers can use the same APIs as those of traditional DL libraries (e.g., TensorFlow~\cite{TensorFlow}) to perform DL inference. In addition, \framework~provides simple interfaces for developers or users to flexibly specify policies such as the latency requirement and energy consumption preferences. Overall, \framework~is readily deployable to provide immediate benefits for wearable applications.


We have implemented the \framework prototype on Android OS (for handheld) and Android Wear OS (for wearable). We evaluated our prototype on COTS smartphones and smartwatches using the state-of-the-art DL models. %
\framework~can effectively identify the optimal partition for offloading under various combinations of device hardware, system configurations, and usage contexts, with the accuracy being up to \textbf{97.9\%}.
\framework~brings on average \textbf{1.95X} and \textbf{2.62X} (up to \textbf{5.08X} and \textbf{23.0X}) DL inference speedup compared to the handheld-only and wearable-only execution strategies, respectively.
In addition, it brings on average \textbf{18.0\%} and \textbf{32.7\%} (up to \textbf{53.5\%} and \textbf{85.5\%}) energy saving compared to the two strategies respectively.
Meanwhile, \framework can adapt its offloading strategies to diverse contexts such as the battery level on either wearable or handheld, the Bluetooth bandwidth, the handheld processor load level, and the user-specified latency requirement. In addition, our pipelining technique helps improve the processing throughput by up to \textbf{84\%} for streaming data compared to the non-pipelining approach.
Finally, \framework incurs negligible runtime and energy overhead.

It should be noted that, there have been various code offloading efforts, including MAUI~\cite{MAUI}, CloneCloud~\cite{CloneCloud}, COMET~\cite{COMET}, DPartner~\cite{Zhang:OOPSLA2012}, and so on. These systems focus on optimizing the general-purpose computation-intensive tasks instead of deep learning applications, and the offloading decisions are often manually defined at the design time (e.g., profiling~\cite{CloneCloud} or manually labeled~\cite{Zhang:OOPSLA2012}). However, \framework intuitively differs from these systems as it relies on the domain knowledge of deep learning models, i.e., the  \textit{data topology} between layers of a DL model rather than the code-level characteristics, and the offloading decision is dynamically made at runtime rather than manually pre-defined.
Additionally, compared to recent efforts on mobile DL offloading such as~\cite{conf/asplos/KangHGRMMT17}, our work specifically focuses on wearable devices, with additional effective mechanisms such as streamed data processing. To summarize, we make the following major technical contributions in this paper.
%

\revise{
\begin{itemize}
\item We conduct to the best of our knowledge the most comprehensive characterization of wearable DL offloading, by applying 8 representatively popular DL models on COTS wearables/smartphones under various settings and quantifying several key tradeoffs.
We demonstrate that whether and how much users can benefit from the wearable-to-handheld offloading depends on multiple factors such as hardware specifications, model structures, etc. We reveal that in some cases partitioning the DL models into two parts and running them separately on the wearable and the handheld would have better performance and quality of user experience.

\item We design and implement \framework, a DL framework for wearable devices.
It intelligently, transparently,
and adaptively offloads DL tasks from a wearable
to a paired handheld.
With the help from local offloading, \framework better preserves users' privacy and thus realizes a more ubiquitous offloading without requiring the Internet connectivity.
    %
%
\framework introduces various innovative and effective techniques such as context-aware offloading, strategic model partition, and pipelining support, to better utilize the processing capacity from nearby handhelds while judiciously managing both the devices' resource and energy utilization.

\item We comprehensively evaluate the \framework approach over COTS wearable and handheld devices. The results demonstrate that \framework can accurately identify the optimal partition strategy, and
strike a much better tradeoff among the end-to-end latency and the energy consumption on both the handheld and the wearable, compared to the wearable-only and the handheld-only strategies.
\end{itemize}
}

The remainder of the paper is organized as follows.
We survey the related work in Section~\ref{sec:related}.
We present our measurements about wearables DL in Section~\ref{sec:back}.
We describe the design and implementation of \framework in Section~\ref{sec:design} and Section~\ref{sec:implementation}, respectively.
We comprehensively evaluate \framework in Section~\ref{sec:eval}.
We discuss the limitations and possible future work in Section~\ref{sec:discuss} and conclude the paper in Section~\ref{sec:conclusion}.

\mysection{Related Work}\label{sec:related}
In this section, we discuss existing literature studies that relate to our work presented in this paper.

\subsection{Ubiquitous Deep Learning}
\revise{
In the past few years, DL is the state-of-the-art AI technique that has been widely applied in numerous domains, such as computer vision, pattern recognition, natural language processing, and so on~\cite{journals/chinaf/WangSLDZ17,journals/chinaf/LiLZHPY016,journals/chinaf/QuWFZY17}.
%
A DL model is essentially a directed graph where each node represents a processing unit that applies certain operations to its input and generates output.
%
Accordingly, developers need to first construct a specific model graph, and then use data to train the model (known as the training stage). Once trained, the model can be applied for prediction (known as the inference stage).

Tremendous efforts have been made towards reducing the computation overhead of DL tasks, making it feasible on resource-constrained devices such as smartphones.
For example, some recent efforts~\cite{conf/huc/LaneGQ15,conf/icassp/ChenPH14,conf/icassp/VarianiLMMG14,MobileNet} have proposed lightweight DL models that can run directly on low-end mobile processors.
Some other efforts such as~\cite{conf/asplos/ChenDSWWCT14,conf/fpga/ZhangLSGXC15,conf/isca/ChenES16,conf/isca/HanLMPPHD16} aimed at building customized hardware accelerators for DL or other machine learning tasks.
Besides, various model compression techniques~\cite{conf/mobisys/LiuLZNLD18,conf/ipsn/LaneBGFJQK16,conf/cvpr/WuLWHC16,conf/nips/DentonZBLF14,conf/mobisys/HanSPAWK16,xu2018deepcache} have been proposed for accelerating the DL task and reducing its energy consumption.

In contrast, \framework~specifically focuses on wearable devices that have specific features and application contexts compared to smartphones. \framework~proposes novel techniques such as strategic model partition and pipelining to efficiently utilize the processing capacity from a nearby handheld.
}

\subsection{Offloading}

Many prior efforts, such as MAUI~\cite{MAUI}, CloneCloud~\cite{CloneCloud} COMET~\cite{COMET}, and DPartner~\cite{Zhang:OOPSLA2012}, have already studied the offloading problem from mobile devices to the remote server or cloud.
In addition, \framework~also learns lessons from the recent work on ``edge cloud'' or ``cloudlet'' offloading~\cite{conf/icpp/YangCWW17,conf/IEEEcloud/YangLCSW17,conf/apsys/HuGHWACPS16}.
All these frameworks are control-centric, as they make decisions at the level of code or function. For example, COMET~\cite{COMET} offloads a thread when the execution time exceeds a pre-defined threshold, ignoring any other information, e.g., considerable data volume (of distributed shared memory) to transfer, wireless network available, etc. CloneCloud~\cite{CloneCloud} makes similar offloading decisions for all invocations of the same function. MAUI~\cite{MAUI} designs an enhanced offloading decision mechanism that makes predictions for every single function invocation separately and considers the entire application when choosing which function to offload. DPartner~\cite{Zhang:OOPSLA2012} requires offline profiling to identify the computation-intensive functions and the programmers' manual efforts to annotate whether these functions are ``\textit{offloadable}''.

However, these \textit{general-purpose} offloading efforts are not sufficiently adequate to the partition decisions of DL model, which essentially depends on the data topology. As a result, layers of a given type within the DL model can have significantly different computational and data characteristics~\cite{conf/asplos/KangHGRMMT17}, and can vary a lot even when executing the same code or functions. In contrast, \framework~differs from these approaches in that its offloading leverages the domain knowledge of DL to make the partition decision. Although \framework still requires the offline profiling, it does not introduce any additional manual efforts (such as annotation) to programmers, and the partition is performed dynamically based on the runtime DL topology. Another important difference of \framework is the careful considerations of practical deployability. That is, existing function-level code offloading approaches are too complex and heavyweight, e.g., the complicated program state synchronization for DSM in COMET~\cite{COMET}, and the high-volume data transfer of DPartner~\cite{Zhang:OOPSLA2012}. Such overhead can deter the deployability on  wearables. \framework designs a very lightweight framework that is easy to be deployed on wearables. Additionally, \framework achieves the satisfactory performance and accuracy when making offloading decisions, which has not been well studied in existing offloading solutions.

\subsection{DL Model Partitioning}
The recently proposed DeepX~\cite{conf/ipsn/LaneBGFJQK16} also partitions DL models for low-power DL inference.
However, DeepX only distributes partitioned submodels onto different \emph{local} processors while \framework
performs collaborative DL inference on two devices and thus needs to
take into consideration the data transfer overhead and many other external factors
that play important roles in making offloading decisions.
Also, DeepX targets at only linear DNN models, while \framework~can handle complex DL models with non-linear structures.
Some other work~\cite{conf/asplos/KangHGRMMT17,journals/corr/OssiaSTRLH17} also split DL computation between client devices and remote clouds.
\framework~instead focuses on the collaboration between wearables and their paired handheld devices in order to preserve the privacy and realize ubiquitous DL without requiring Internet connectivity.
Several unique challenges thus stem from the architecture we have chosen, such as balancing the resource consumption on both mobile devices.
Furthermore, \framework~introduces optimizations for streaming data processing, a missing feature in prior work.


\mysection{A measurement Study of Wearable DL}\label{sec:back}

In this section, we begin with some empirical studies to demystify the performance and limitations of running DL tasks on wearables.

\begin{table}[t]
	\centering
	\scriptsize
 	\begin{tabular}{|l|l|l|l|}
 	\hline
 	\multicolumn{1}{|c|}{\textbf{Model}} & \multicolumn{1}{c|}{\textbf{App}} & \multicolumn{1}{c|}{\textbf{Input}} & \multicolumn{1}{c|}{\mrev{\textbf{FLOPs}}}\\\hline
 	\emph{MNIST}~\cite{MNIST} & digit recognition & grayscale image & \mrev{15M}\\\hline
 	\emph{MobileNet}~\cite{MobileNet} & image classification & rgb image & \mrev{580M}\\\hline
 	\emph{GoogLeNet}~\cite{Inception} & image classification & rgb image & \mrev{2G}\\\hline
 	\emph{LSTM-HAR}~\cite{LSTM-HAR} & activity recognition & mobile sensor & \mrev{180M}\\\hline
 	\emph{DeepSense}~\cite{conf/www/YaoHZZA17} & activity recognition & mobile sensor & \mrev{550M}\\\hline
 	\emph{TextRNN}~\cite{TextRNN} & document classification & word vectors & \mrev{11M}\\\hline
 	\emph{DeepEar}~\cite{conf/huc/LaneGQ15} & emotion recognition & raw sound & \mrev{9M}\\\hline
 	\emph{WaveNet}~\cite{WaveNet} & speech recognition & mfcc features & \mrev{3.8G}\\\hline
	\end{tabular}
\caption{8 deep learning models used in this work.}\label{tab:models}
\end{table}

%
%
In this work, we study 8 state-of-the-art DL models that have been widely adopted in various applications, as shown in Table~\ref{tab:models}.
For the \emph{LSTM-HAR} model~\cite{LSTM-HAR}, we use a popular configuration as 2-layer stacked, 1024 hidden state size to carry out our experiments.
For other models, we use the default configurations as described in the original literature or open-sourced repositories.
These models range from natural language processing, audio processing, to computer vision tasks and mobile sensor intelligence, all of which are well suited to ubiquitous and wearable scenarios.
\mrev{The rightmost column in Table~\ref{tab:models} also lists the number of FLOPs (floating point operations) for conducting a single inference for each model.}
%
%
It is worth mentioning that DL models are often generalized and can be used in many different tasks with very few customization efforts. For example, the \emph{LSTM} model used for language modeling can also be applied to problems such as machine translation~\cite{conf/nips/SutskeverVL14}, question answering~\cite{conf/acl/WangN15}, and handwriting generation~\cite{journals/corr/Graves13}.
In particular, \framework does not assume any specific DL model structure, and can work with all of them.

We envision that DL will become an essential part in the wearable ecosystem due to wearables' unique sensing capabilities.
However, running computation-intensive DL tasks on wearable devices is quite challenging due to wearables' relatively limited processing capabilities.
A possible approach is thus to offload the workload from a wearable to its paired handheld.
We choose the handheld over the cloud because offloading to the handheld does not require the Internet connectivity that can incur  high energy and monetary cost. Doing so also minimizes risks of privacy breach because the potentially sensitive data is never leaked to the Internet.
Note that there are quite a lot prior work~\cite{conf/globecom/0004YZ15,conf/mm/ShiYHH15,conf/mobisys/KoLC16} targeting at wearable offloading for better performance (see Section~\ref{sec:related}).
However, none of them studies DL tasks, thus leaving an important question unanswered:
\textit{whether and how much offloading to a handheld can benefit DL on wearables?}
To answer this question, we carry out a set of experiments on 8 popular DL models and various hardware setups.
Our experiment results show that whether and how much users can benefit from offloading depends on multiple factors.
In particular, we will reveal that in some cases partitioning the DL models into two parts and run them separately on the wearable and the handheld would be a more promising option. We call such a scheme ``\emph{partial offloading}''.

\begin{table}[t]
	\centering
	\scriptsize
 	\begin{tabular}{|l|l|l|l|l|}
 	\hline
 	\multicolumn{1}{|c|}{\textbf{Device}} & \multicolumn{1}{c|}{\textbf{CPU}} & \multicolumn{1}{c|}{\textbf{Memory}} & \multicolumn{1}{c|}{\textbf{GPU}}\\\hline
 	Nexus 6 & Quad-core Krait 450 & 3 GB RAM & Adreno 420\\\hline
 	LG Urbane & Quad-core Cortex-A7 & 512MB RAM & Adreno 305\\\hline
 	Galaxy S2 & Dual-core Cortex-A9 & 1GB RAM & Mali-400MP4\\\hline
 	\end{tabular}
\caption{Hardware specifications for wearables and smartphones used in this work.}\label{tab:hardware}
\end{table}

\begin{figure*}
\centering
\scriptsize
\subfloat[\emph{GoogLeNet} model.]{\includegraphics[width=0.25\textwidth]{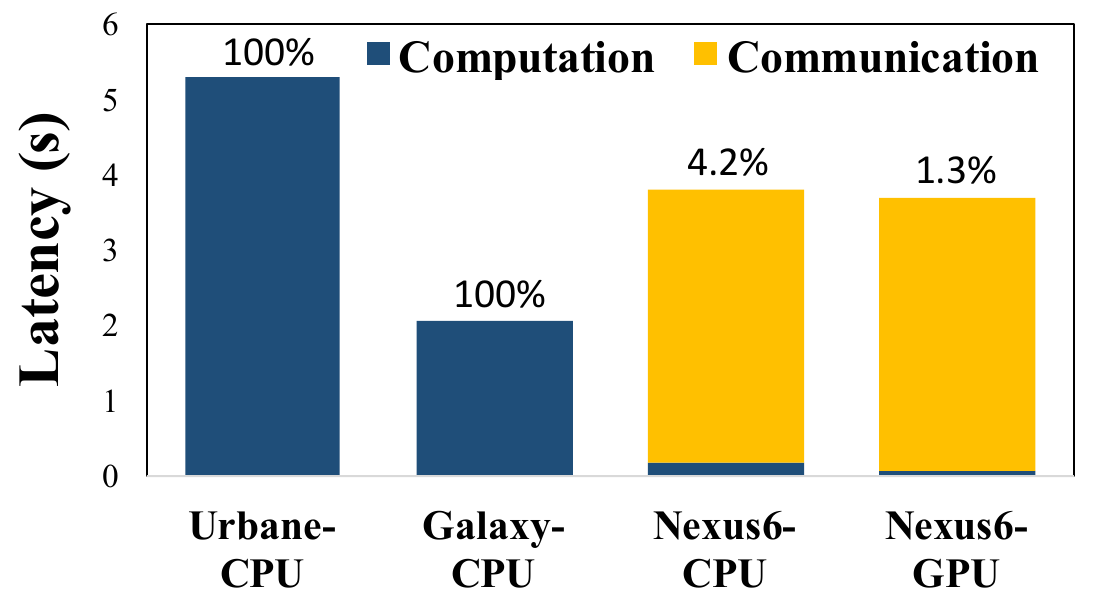}}
\subfloat[\emph{LSTM-HAR} model.]{\includegraphics[width=0.25\textwidth]{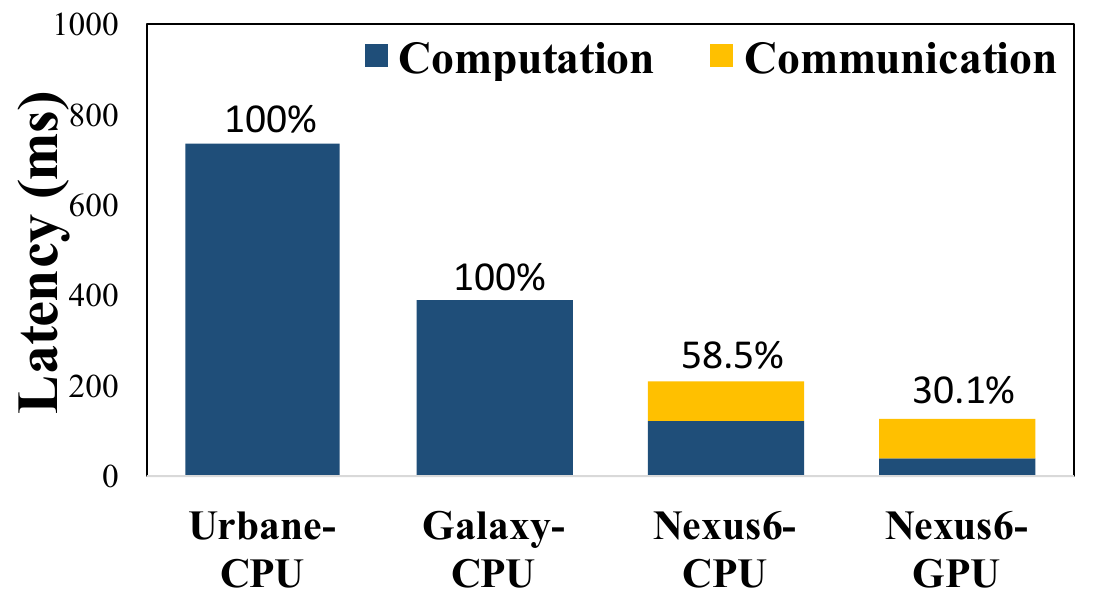}}
\subfloat[\emph{MNIST} model.]{\includegraphics[width=0.25\textwidth]{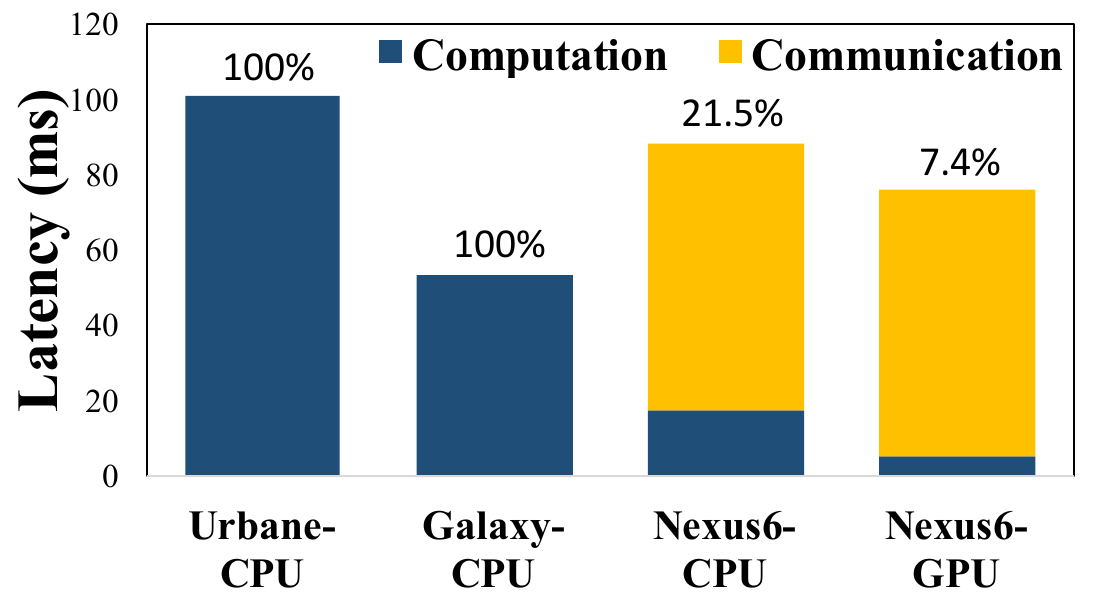}}
\subfloat[\emph{DeepSense} model.]{\includegraphics[width=0.25\textwidth]{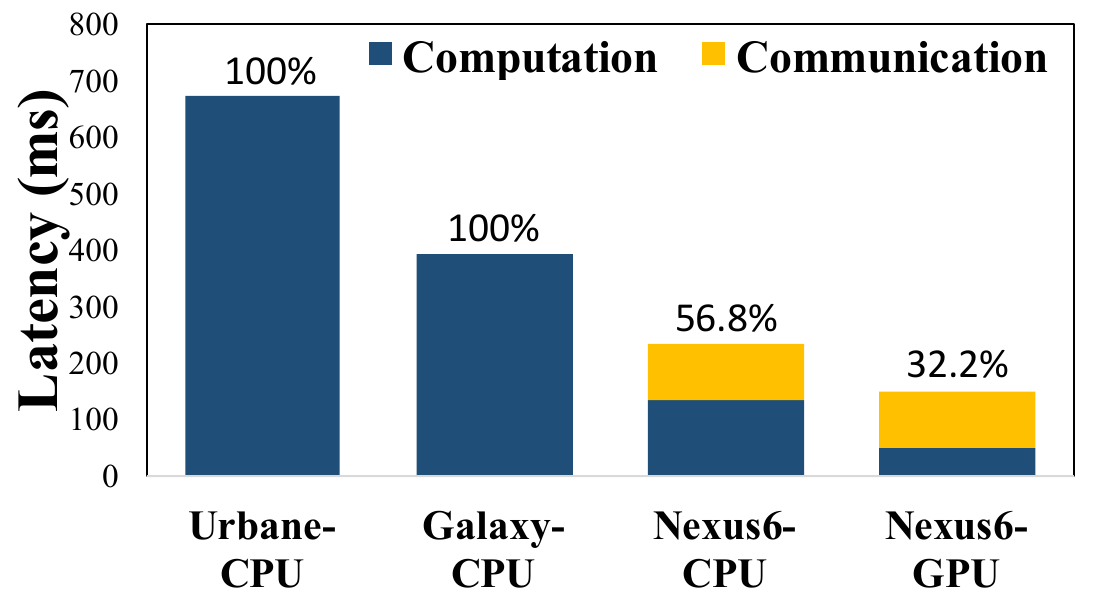}}
\caption{End-to-end latency breakdown for different models and offloading scenarios.
The upper percentage indicates the proportion of computation time among the overall latency.
Offloading to the handheld is often slower than wearable execution due to the high data transfer delay via Bluetooth.
}
\label{fig:latency_breakdown}
\end{figure*}

\begin{figure*}
\centering
\scriptsize
\subfloat[\emph{MobileNet} model.]{\includegraphics[width=0.25\textwidth]{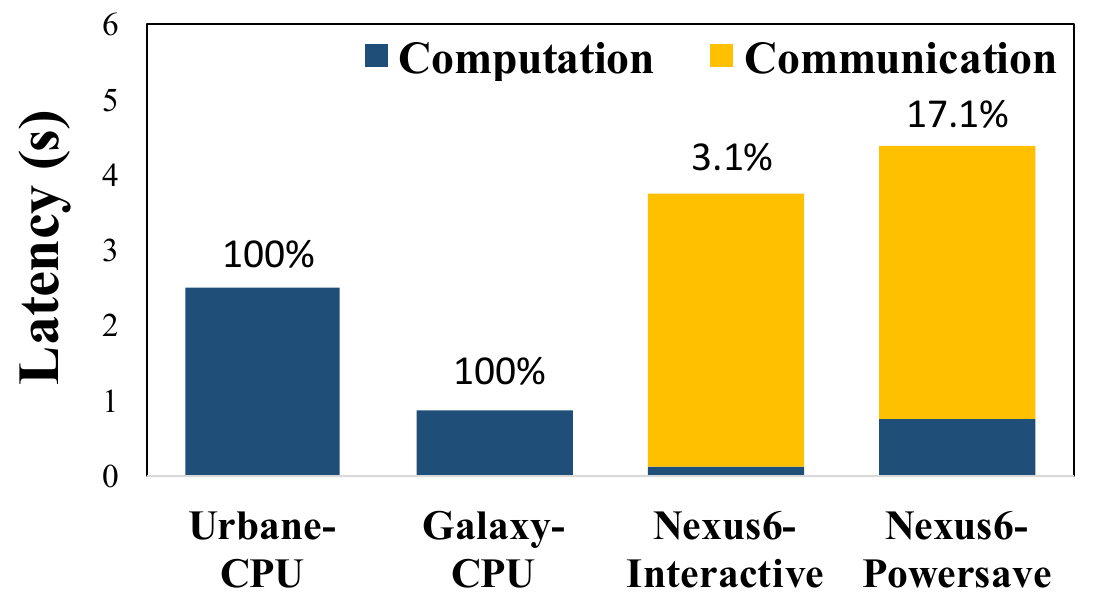}}
\subfloat[\emph{TextRNN} model.]{\includegraphics[width=0.25\textwidth]{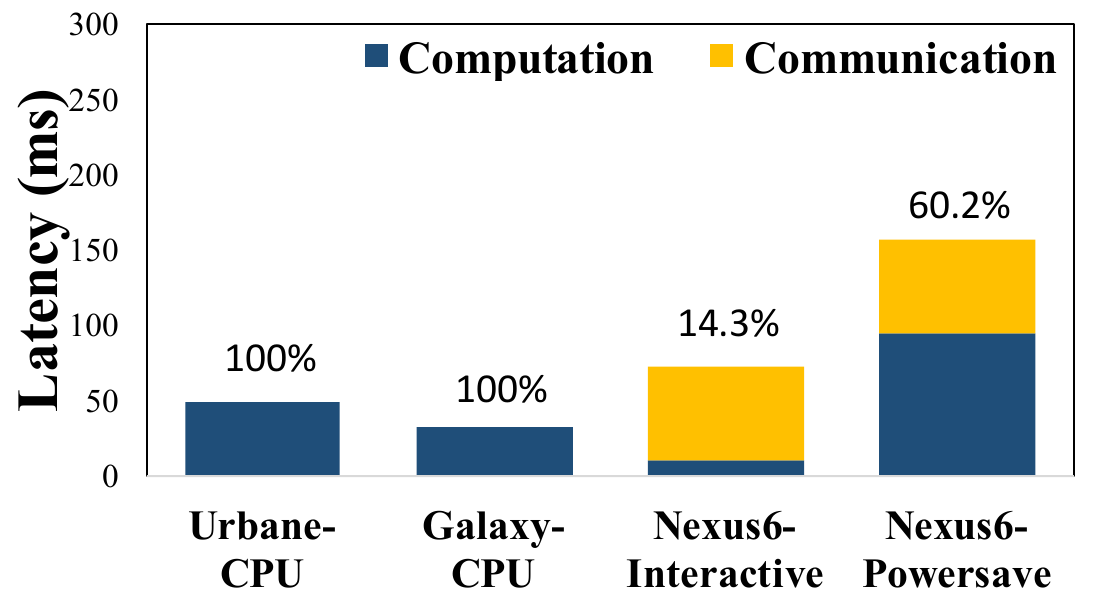}}
\subfloat[\emph{DeepEar} model.]{\includegraphics[width=0.25\textwidth]{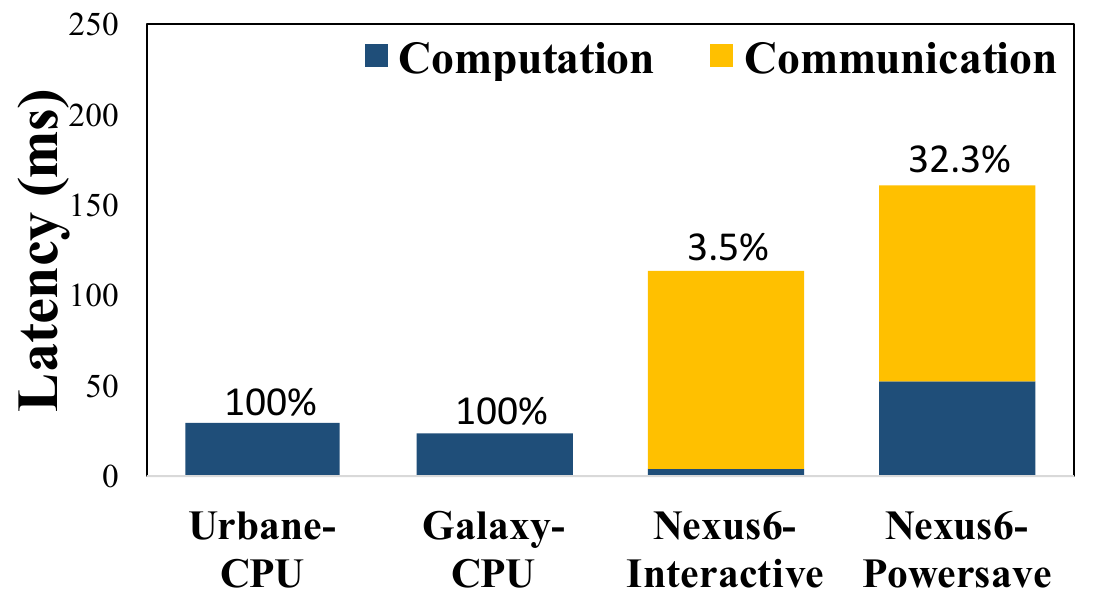}}
\subfloat[\emph{WaveNet} model.]{\includegraphics[width=0.25\textwidth]{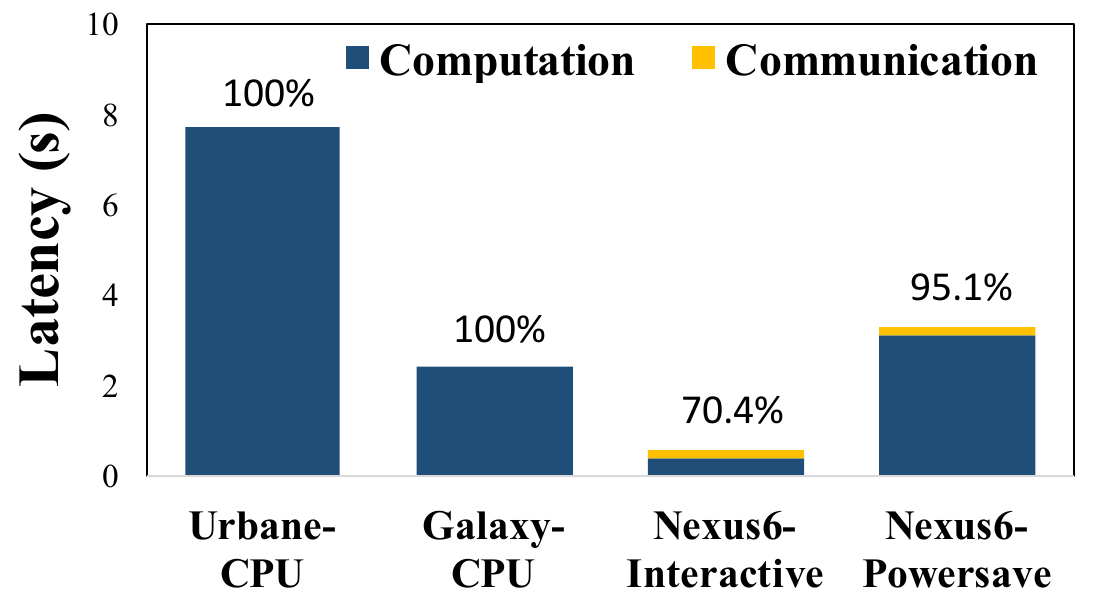}}
\caption{End-to-end latency breakdown under different handheld CPU governor.
The upper percentage indicates the proportion of computation time among the overall latency.
The device status such as current CPU governor can have key impacts on making choice about offloading.
}
\label{fig:latency_breakdown_powersave}
\end{figure*}

\textbf{Experimental setup.}
We use a Nexus 6 smartphone running Android 7.0 as the handheld device, and an LG Watch Urbane as the (real) wearable device.
We also use an old phone, Galaxy S2 released in 2011, to emulate head-mount devices such as Vuzix M1000~\cite{M1000} that shares hardware similar to Galaxy S2.
Table~\ref{tab:hardware} elaborates the hardware specifications of these three devices used in this study.
We use TensorFlow~\cite{TensorFlow} and an open-source library RSTensorFlow~\cite{rstensorflow} to support running DL tasks on mobile CPU and GPU.
We use Bluetooth for the data transfer between wearable and handheld due to Bluetooth's wide availability on wearable and its energy efficiency.
\footnote{Also Google has recommended it as the proper way of performing data communication on wearable devices~\cite{DataLayer}.}
For energy measurement, we build the power model for the smartphone by using the Monsoon Power Meter~\cite{Monsoon} (following a high-level approach of component-based power modeling~\cite{conf/codes/ZhangTQWDMY10}),
or obtain the model from the literature~\cite{conf/mobisys/LiuCQGLWC17} for smartwatch.
All experiments are carried out by fixing the distance between the wearable and handheld (0.5m) unless otherwise stated.

\begin{figure*}
\centering
\scriptsize
\subfloat[\emph{GoogLeNet} model.]{\includegraphics[width=0.4\textwidth]{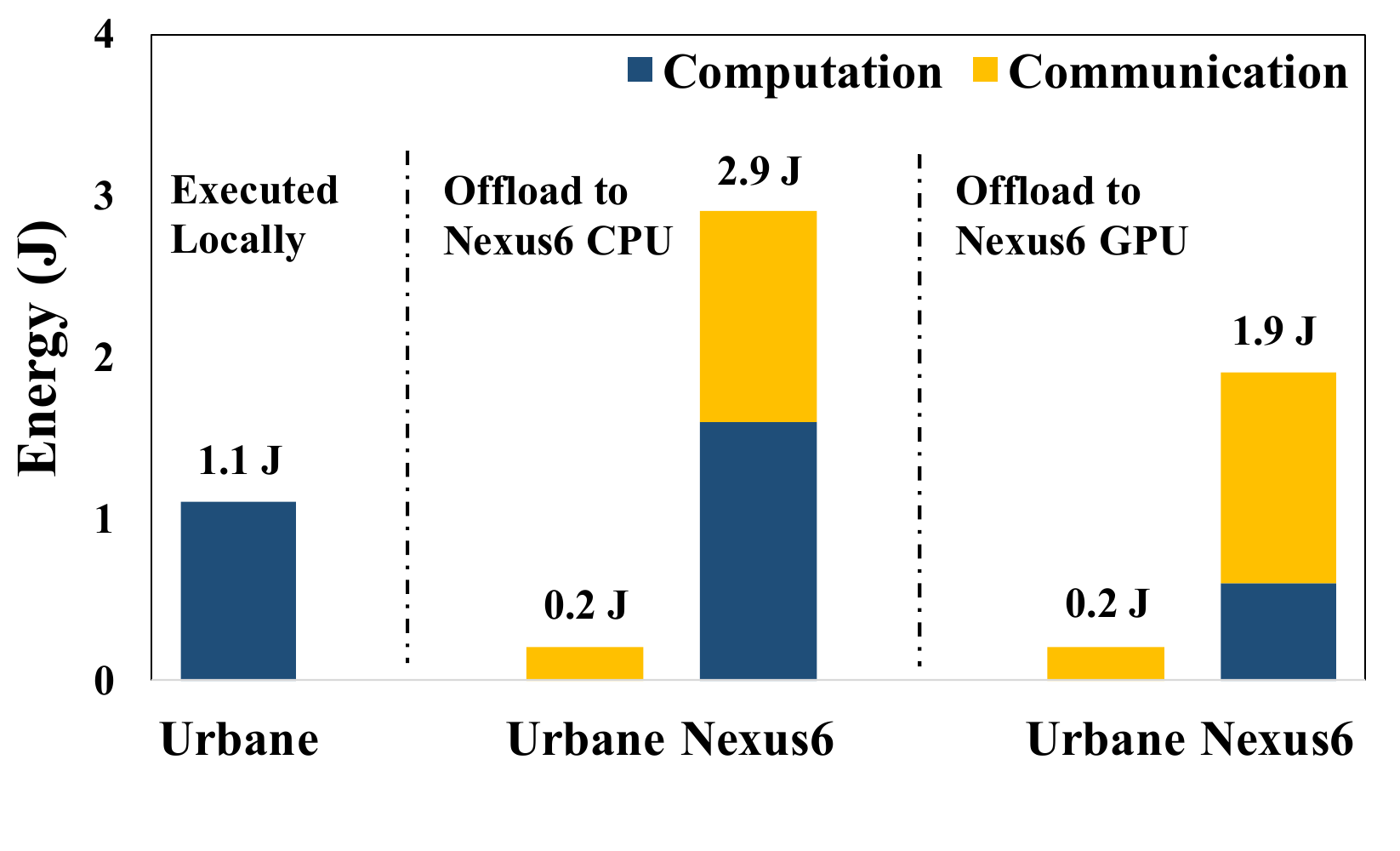}}
\subfloat[\emph{LSTM-HAR} model.]{\includegraphics[width=0.4\textwidth]{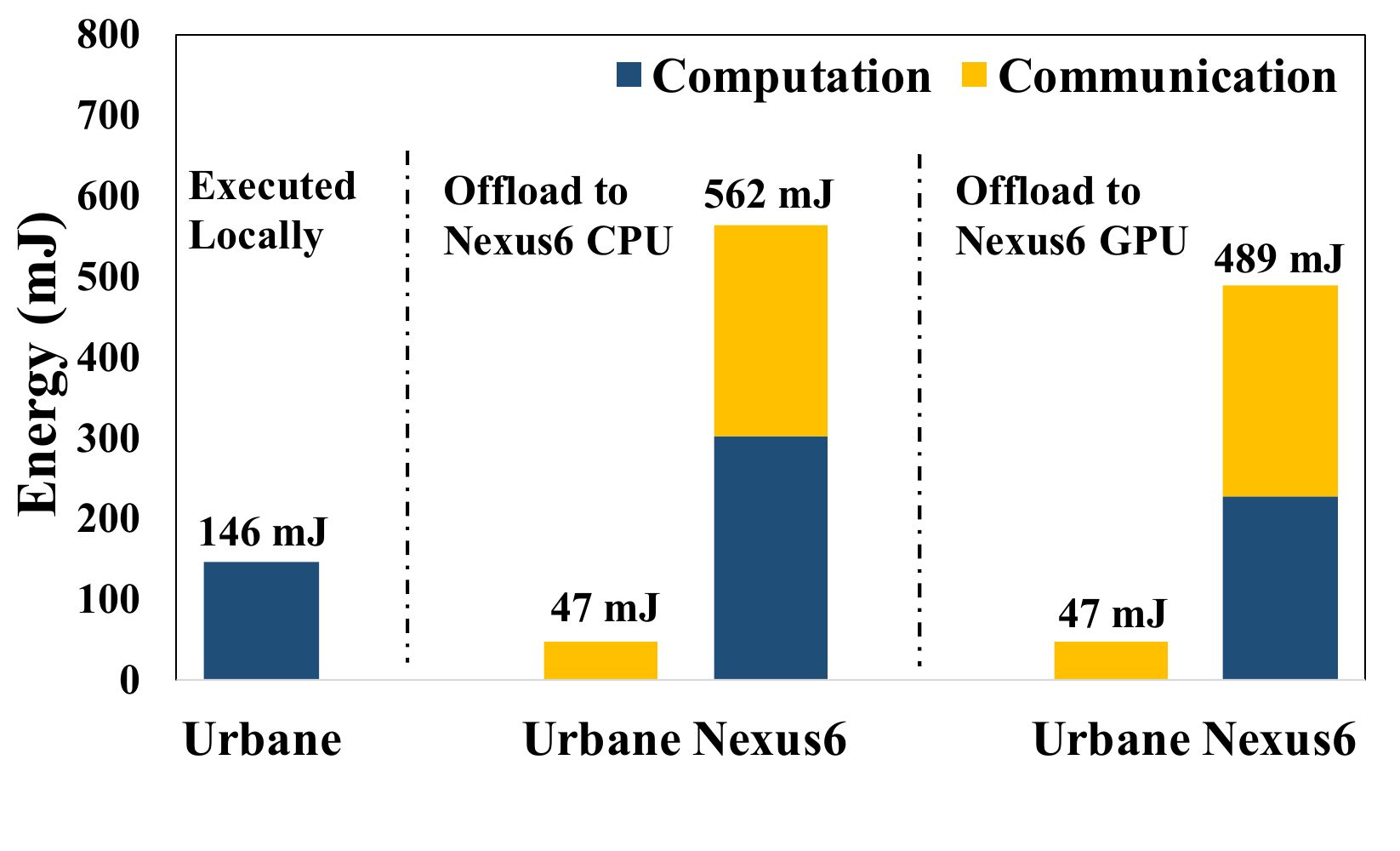}}
\caption{Energy breakdown on both wearable and handheld devices for different running strategy.
Offloading to the handheld can sometimes consume more energy than wearable execution due to the high energy overhead consumed by Bluetooth module.
}
\label{fig:energy_breakdown}
\end{figure*}

\textbf{Factors of determining offloading decisions.}
Figure~\ref{fig:latency_breakdown} and~\ref{fig:latency_breakdown_powersave} indicate the latency breakdown of four popular DL models under different offloading scenarios.
In each plot, the two left columns present the latency of executing the whole model on different wearable devices (LG Urbane and S2), while the two right columns show the latency of offloading them to handheld processors (CPU and GPU respectively).
The percentage indicates the proportion of computation time (as opposed to network delay) within the overall latency.
%
Our key observation is that \emph{although offloading to handheld CPU and GPU can dramatically reduce the computation time, e.g., more than 10 times for the \emph{GoogLeNet} model, the end-to-end latency is often not reduced due to the high data transfer latency over Bluetooth}.
The results indicate that making a judicious offloading decision can have significant impacts on the user experience.
For example, running the \emph{DeepEar} model locally on LG Urbane can reduce up to 74\% of latency compared to running it on handheld CPU, while for the \emph{DeepSense} model, running it locally leads to more delay compared to offloading to a handheld.

Overall, the optimal decision depends on various factors described below.

(1) \textbf{Device heterogeneity.}
There exist diverse wearable devices with highly heterogeneous hardware, ranging from a tiny smart ring to a large head-mount device for virtual reality. 
For example,
our experiments show that for LG Urbane and Galaxy S2, they often need to adopt different offloading strategies: to achieve the lowest latency for the \emph{GoogLeNet} model, LG Urbane should offload the task to Nexus 6 while Galaxy S2 does not need to do so according to Figure~\ref{fig:latency_breakdown}(a).

(2) \textbf{Model structure.}
Different DL models can vary a lot in terms of computational overhead and input size.
Models with high a computational overhead and a small input size such as \emph{DeepSense} and \emph{WaveNet} are more adequate for being offloaded to handhelds, while other models may not benefit from offloading such as \emph{DeepEar}.

(3) \textbf{Processor status.}
In real-world application scenarios, handheld CPUs often run under different governors adapting to different device environments, e.g., switching from the default \textit{interactive} governor (high frequency) to the \textit{powersave} governor (low frequency) when the screen is turned off or the battery level is low.
Observed from Figure~\ref{fig:latency_breakdown_powersave}, CPU status can have substantial impacts on the latency as well as the offloading strategy.
Take \emph{WaveNet} as an example. It takes almost 7X more time under the \textit{powersave} governor than the \textit{interactive} governor, with the former rendering offloading no longer beneficial.
While enforcing the handheld to switch to a high-power governor is sometimes possible, there are other scenarios where the handheld CPU/GPU is inherently overloaded (e.g., by other computationally intensive apps that are running concurrently).

(4) \textbf{Latency vs. energy preference.}
Besides the end-to-end latency, the energy consumption is another key metric to consider as wearable devices have small battery capacities~\cite{conf/mobisys/LiuCQGLWC17}.
As shown in Figure~\ref{fig:energy_breakdown}, although offloading can help save wearable battery, it will also cause the non-trivial energy consumption for the handheld (around 2.9 J for Nexus 6 CPU for \emph{GoogLeNet}).

Overall, the above results indicate the challenge of \emph{balancing the tradeoff among three factors
when making judicious offloading decisions: end-to-end latency, energy consumption of the wearable, and energy consumption of the handheld}.
In real-world scenarios, a static policy may not always satisfy users' requirements.
For instance, when a user's handheld (wearable) is low on battery, \framework needs to focus on saving the energy for the handheld (wearable). Therefore it is necessarily beneficial to adjust the offloading decisions dynamically based on external factors such as battery life, network condition, and CPU/GPU workload.

\begin{figure}[t]
	\centering
	\includegraphics[width=0.49\textwidth]{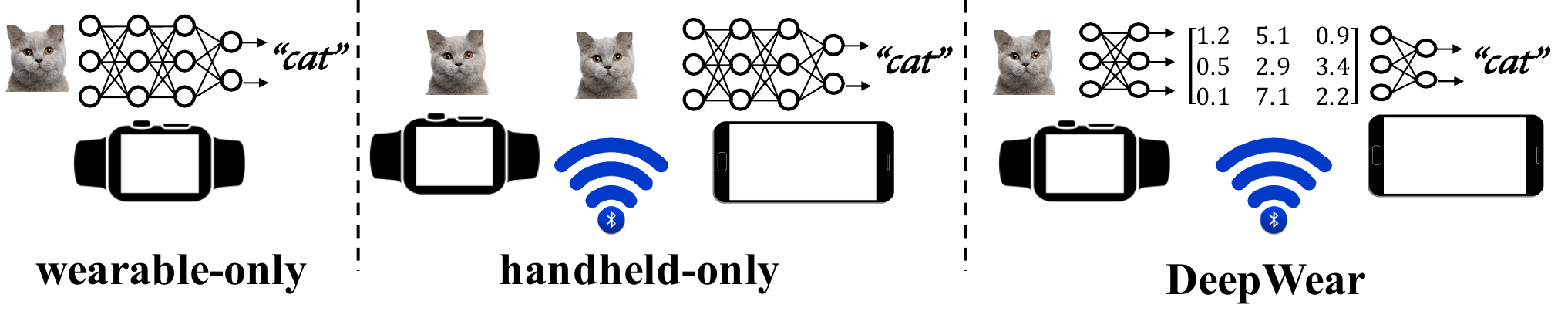}
	\caption{Different wearable DL execution approaches.
	wearable-only (offload nothing), handheld-only (offload everything), and DeepWear (partial offloading). Offloading nothing means executing all DL task on wearable.
	Offloading everything means offloading all DL task to handheld. Partial offloading, which is adopted in \framework, means partitioning computation among wearable and handheld.
	}
	\label{fig:partial_offloading}
\end{figure}

\textbf{Partial offloading.}
The preceding pilot experiments consider only two scenarios: offloading the whole DL model to the handheld or executing it locally.
Our further investigation indicates that partial offloading, i.e.,
dividing the DL model into two sub-models and executing them separately on the wearable and the handheld as shown in Figure~\ref{fig:partial_offloading}, can sometimes achieve even better results.

\begin{figure}[t]
	\centering
	\includegraphics[width=0.49\textwidth]{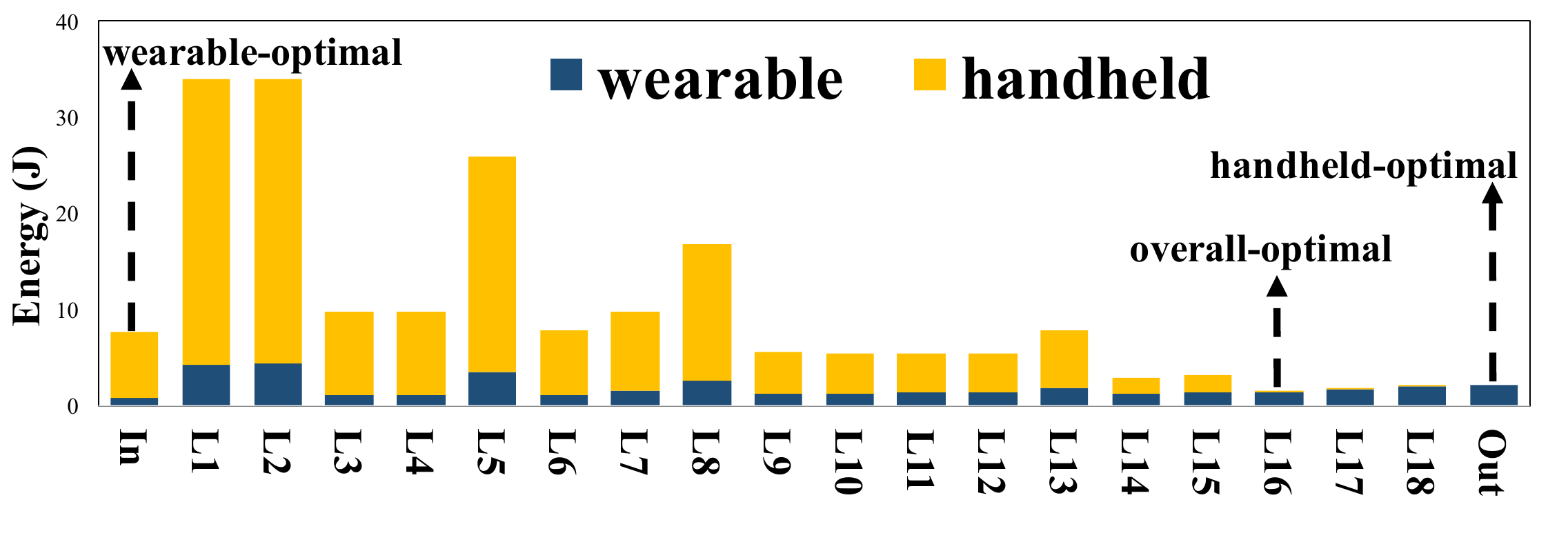}
	\caption{Energy consumption of running \emph{GoogLeNet} on LG Urbane and Nexus 6 with different partition points.
	We only select 20 partition points to present the figure.
	X-axis presents the layers that we select as partition point, after which output data is sent to handheld for continuous processing.
	The left-most bar represents handheld-only processing and the right-most bar represents wearable-only processing.}
	\label{fig:inception_partition}
\end{figure}

We confirm the benefit of partial offloading through controlled experiments.
Figure~\ref{fig:inception_partition} plots the energy consumption with different partition points for the \emph{GoogLeNet} model.
The X-axis presents the layer that we select as partition point, after which the output data is sent to handheld for further processing.
The left-most and right-most bars correspond to handheld-only and wearable-only processing, respectively.
\mrev{Note that the energy consumption of the handheld in Figure~\ref{fig:inception_partition} (and all energy results thereafter) is calibrated as $E = original\_E / Handheld\_capacity * Wearable\_capacity$. $original\_E$ is the absolute energy consumed by the handheld; $Handheld\_capacity$ and $Wearable\_capacity$ are the battery capacity of the handheld (3220 mAh for Nexus 6) and the wearable (410 mAh for LG Urbane), respectively.
Since the phone and watch have different battery capacities, the above adjustment essentially calibrates the phone and wearable's energy consumption with respect to their heterogeneous actual battery capacities.
}

As shown in Figure~\ref{fig:inception_partition}, executing the model locally without offloading is the most energy-efficient for the handheld, while offloading the whole task to the handheld consumes the least amount of energy for the wearable.
However, users often care about the battery life of both devices, therefore we need to find an optimal partition to achieve the least total energy consumption.
In this case, the overall optimal partition point resides in an internal layer (L16).
Doing such a partial offloading helps save around 84\% and 29\% of energy compared to the wearable-only and handheld-only strategies, respectively.

\begin{figure}[t]
	\centering
	\includegraphics[width=0.49\textwidth]{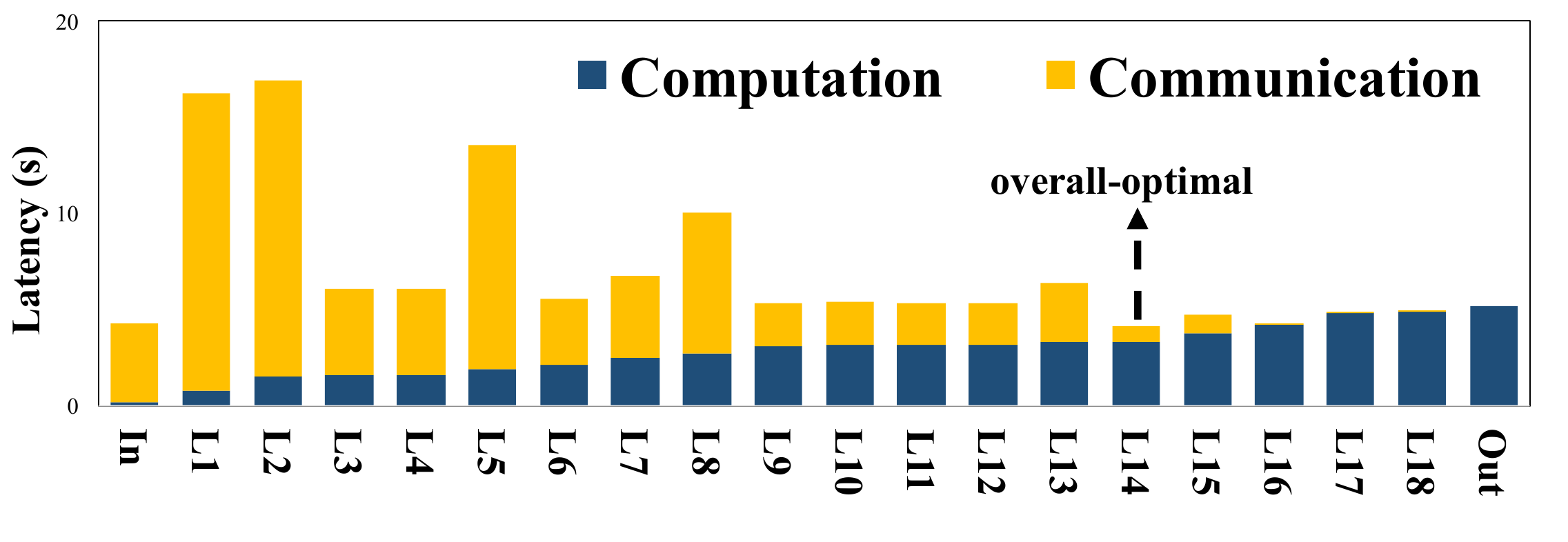}
	\caption{\mrev{End-to-end latency of running \emph{GoogLeNet} on LG Urbane and Nexus 6 with different partition points. The experimental setup is the same as that in Figure~\ref{fig:inception_partition}.}}
	\label{fig:inception_partition_latency}
\end{figure}

\mrev{Using the same setup as that in Figure~\ref{fig:inception_partition}, Figure~\ref{fig:inception_partition_latency}
plots the end-to-end latency with different partition points for the \emph{GoogLeNet} model.
As shown, performing partial offloading may also help minimize the overall latency (L14 in  Figure~\ref{fig:inception_partition_latency}).} This is because an internal layer may yield a small intermediate output compared to the original input size, thus reducing the network transmission delay.
Therefore, a key design decision we make for \framework is to support partial offloading.

\mysection{The \framework Design}\label{sec:design}

Our measurements in Section~\ref{sec:back} indicate that it is challenging to develop an offloading framework for wearables with various factors being considered.
%
We thus argue that flexible and efficient DL offloading support should be provided as a ready-made service to all applications, as opposed to being handled by app developers in an ad-hoc manner.
To this end, we propose a holistic framework called \framework, which can help applications optimally determine \emph{whether or not, how, and what to offload}.
We now describe the design details of \framework whose design goals include the following.



\noindent $\bullet$ \textbf{Latency-aware.}
Different DL apps have diverse latency requirements, ranging from dozens of milliseconds (augmented reality) to several minutes (activity tracking).
As a result, \framework~should meet the appropriate user-perceived latency requirement, which is given by app developers, as the foremost goal to satisfy.

\noindent $\bullet$ \textbf{Working with off-the-shelf DL Models.}
\framework~should not require developers' additional efforts to retrain the deep learning models.
This is important as most app developers today utilize only off-the-shelf models in an ``as-it-is'' style.

\noindent $\bullet$ \textbf{No accuracy loss.}
\framework should not compromise the accuracy when running DL models under diverse settings.
In other words, \framework should maintain consistently adequate accuracy results regardless of the offloading decision.

\noindent $\bullet$ \textbf{Trade-off flexible.}
\framework~should flexibly balance the tradeoff between the latency and energy based on external factors such as the device battery life on both the wearable and handheld devices.

\noindent $\bullet$ \textbf{Developer-friendly.}
\framework~should provide developers with simple API, as simple as the facilities provided off-the-shelf deep-learning frameworks/libraries such as TensorFlow, Caffe2, PyTorch, etc.
More specifically, \framework~should abstract wearable and handheld devices as one entity by shielding low-level details such as offloading decisions.
For example, developers should be freed from programming on the data transferring via the Android APIs, which is extremely tedious and error-prone.

\mysubsection{Architecture Overview}\label{sec:overall_architecture}

\begin{figure}[t]
	\centering
	\includegraphics[width=0.50\textwidth]{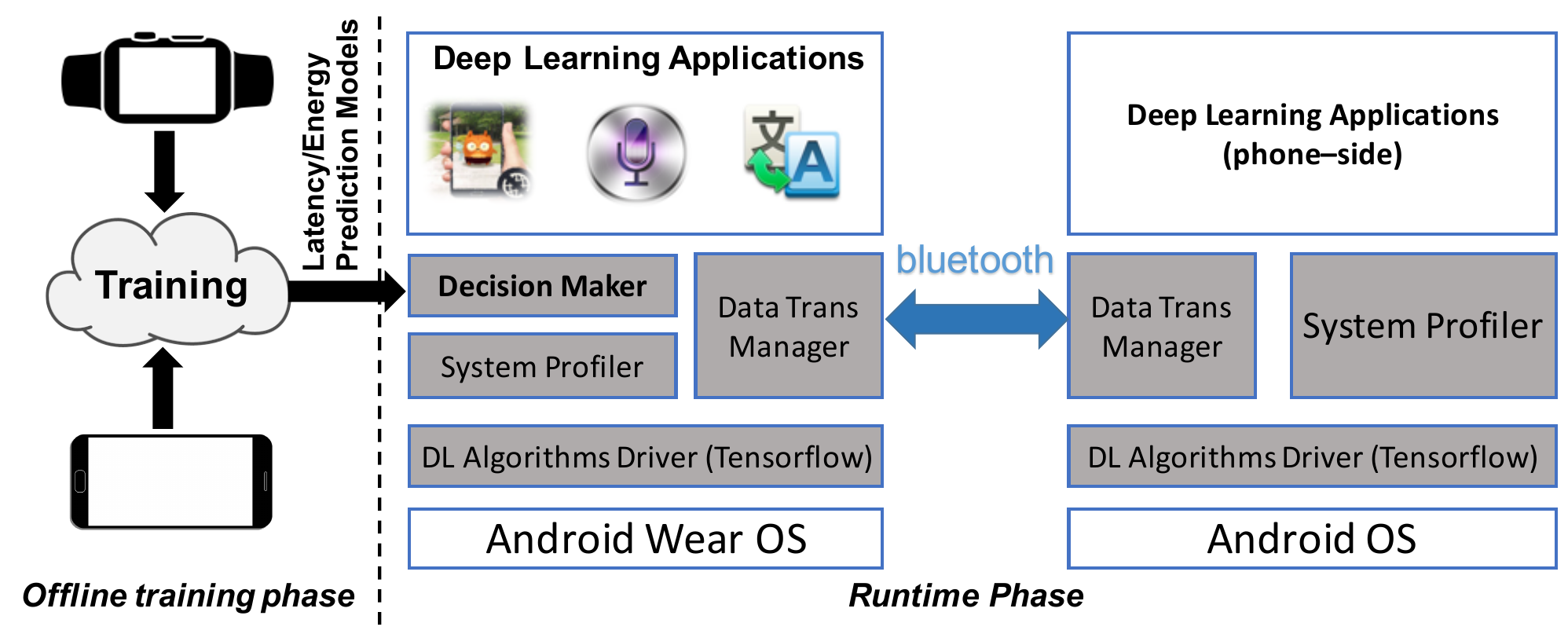}
	\caption{Overview of \framework.
	Grey parts constitute a library provided for deep learning application developers.
	}
	\label{fig:architecture}
\end{figure}

The overall architecture of \framework~is shown in Figure~\ref{fig:architecture}.
To use \framework, there are two main steps involved.

(1) \textbf{The offline training phase} involves a one-time effort of constructing the latency and energy prediction models\footnote{The latency and energy prediction models should be distinguished from the DL models themselves.}, i.e., given a DL model structure, what is the end-to-end latency and energy consumption to run this model on a given device (see Section~\ref{sec:prediction_model}).
\revise{It should be mentioned that, the latency and energy prediction models are device-specific and heavily  depend on the underlying hardware architecture. However, building such a model is a one-time effort and does not incur much engineering overhead based on our experience. In practice, the device or DL software vendors can perform such profiling and let developers download the model. }

(2) \textbf{The runtime phase}, where DL applications rely on \framework to perform adaptive offloading for DL tasks.
There are following major components.

\noindent $\bullet$ \textbf{Decision Maker} is the core part of \framework. Given a DL model to run, it identifies the optimal model partition point based on the latency/energy prediction models and both devices' running status.
The decision dictates which part of the model should be executed locally and which part should be offloaded to the paired handheld, including two special cases of offloading none or the entire task.
A key logic of the Decision Maker is to balance the tradeoff between the latency and the energy consumption (see Section~\ref{sec:decision}).

\noindent $\bullet$ \textbf{System Profiler} periodically profiles the system status such as the pairing state, processor status, and the Bluetooth bandwidth,
which will be used by the Decision Maker to balance key tradeoffs.

\noindent $\bullet$ \textbf{DL Algorithms Driver} is the library that implements the DL algorithms. Currently, \framework directly employs TensorFlow~\cite{TensorFlow} as the driver.

\noindent $\bullet$ \textbf{Data Transmission Manager} deals with the data transmission between the wearable and its paired handheld. It is realized using the standard Data Layer API~\cite{DataLayer} in Android.

\noindent $\bullet$ \textbf{Developer API Wrapper} is the developer interface through which DL applications can be easily developed to use the deep learning libraries with transparent offloading support.
We present the design details in Section~\ref{sec:developer_APIs}. 

\mysubsection{Deriving Prediction Models}\label{sec:prediction_model}



Now we consider how to construct the prediction model of the latency and energy for a given (partial) DL model.
A straightforward way is modeling each layer individually and then combining the prediction models across all layers into the final prediction model.
To demonstrate the feasibility of this approach, we carried out controlled experiments via running DL models and logging the latency/energy in total as well as for each layer.
\footnote{Built-in TensorFlow functionality to log individual layer performance: \url{https://github.com/tensorflow/tensorflow/blob/master/tensorflow/core/util/stat_summarizer.h}}
Through this controlled experiment, we find that to compute the latency/energy consumption of a given (possibly partial) DL model, we can compute the incurred latency/energy for every single layer and then sum them up.
In fact, summing up the latency/energy across all layers yields no more than 1.82\% of deviation compared to the direct measurement, for the eight models shown in Table~\ref{tab:models}.

Nevertheless, we still need to deal with a practical challenge: there exist a large number of layer types inside a DL model (e.g., more than 100 types supported in TensorFlow). As a result, making a prediction model for each of them can incur substantial training overhead.
Fortunately, we find that among those hundreds of layer types, only a small number of them are responsible for typical workloads on wearables: convolutional (conv), fully-connected (fc), pooling, and activation layers.
%
As shown in Table~\ref{tab:pred_model_type}, these four layer types constitute up to more than 90\% of the inference latency of popular DL models.
Although current \framework considers only these layers, other layer types can be easily incorporated into our framework.
\revise{It is quite important to note that for RNN models, a recurrent layer is composed of fully-connected layer and activation layer. Therefore, by modeling the aforementioned layers, i.e., convolutional, fully-connected, pooling, and activation layers, we are able to accommodate the RNN model as well.}
We next describe the methodology of building a prediction model of latency/energy for a given layer.

\textbf{Latency Prediction.}
We observe that even for the same layer type, there might be a large latency variation across different layer parameters (e.g., the kernel sizes of convolution layers).
Thus, we vary the configurable parameters of the layer and measure the latency for each parameter combination.
We use the collected latency data to train and test our prediction models.
As shown in Table~\ref{tab:latency_prediction_result}, we use a combination of decision tree and linear regression to model the latency.
The former is used to classify some types (i.e., convolution, pooling, and activation) into sub-types based on metrics such as the kernel size\footnote{We observe that there are only limited kinds of kernel size used in current CNN models, which is 1X1, 3X3, 5X5, 7X7, and 11X11.} and the activation function.
We then apply a linear-regression model to each of those sub-types to get the final predicted results.
As shown in Table~\ref{tab:latency_prediction_result}, our latency prediction models perform well, especially for the two most computation-intensive layers: convolution and fc, with a high variance score of 0.993 and 0.945, respectively. Here, we use the Coefficient of Determination ($R^2$)~\cite{r2} to measure the accuracy.
\mrev{$R^2$ is a commonly used metric for evaluating regression models.
It assesses how well a model predicts future outcomes.
$R^2$ is calculated as $1-\frac{SSE}{SST}$ where $SSE$ and $SST$ are the sum of squared errors of the regression model and the sum of squared errors of the baseline model (always using the mean as the prediction), respectively.}

\textbf{Energy Prediction.}
We use a similar approach to predicting the energy consumption of a layer.
In our study, we typically build power models for the smartphone by using the Monsoon Power Meter~\cite{Monsoon} (following a high-level approach of component-based power modeling~\cite{conf/codes/ZhangTQWDMY10})
or obtain them from the literature~\cite{conf/mobisys/LiuCQGLWC17} for smartwatch.
All experiments are done with device screen off, and the energy data we used is subtracted by the baseline power in the idle state.

As shown in Table~\ref{tab:latency_prediction_result}, our energy prediction model also has a satisfactory accuracy (\textgreater~92\%) for 3 out of the 4 layer types.
The Pooling layer has a lower accuracy (0.772).
Nevertheless, as shown in Table~\ref{tab:latency_prediction_result} this layer contributes little to the overall latency and energy compared to other layers.

\begin{table}[t]
	\centering
	\scriptsize
 	\begin{tabular}{|l|l|l|l|l|l|}
 	\hline
 	\textbf{Model} & \textbf{Conv} & \textbf{Fc} & \textbf{Pooling} & \textbf{Activation} & \textbf{Total}\\\hline
 	\emph{MNIST} & 39.0\% & 54.2\% & 1.1\% & 3.1\% & 97.4\%\\\hline
 	\emph{MobileNet} & 45.4\% & N.A. & N.A. & 51.1\% & 96.5\%\\\hline
 	\emph{GoogLeNet} & 80.2\% & 0.1\% & 7.5\% & 8.1\% & 95.7\%\\\hline
 	\emph{LSTM-HAR} & N.A. & 8.4\% & N.A. & 87.8\% & 96.2\%\\\hline
 	\emph{DeepSense} & 51.6\% & 21.1\% & N.A. & 25.3\% & 98.0\%\\\hline
 	\emph{TextRNN} & N.A. & 16.0\% & N.A. & 79.1\% & 95.1\%\\\hline
 	\emph{DeepEar} & N.A. & 92.6\% & N.A. & 7.2\% & 99.8\%\\\hline
 	\emph{WaveNet} & 82.6\% & N.A. & N.A. & 11.6\% & 94.1\%\\\hline
	\end{tabular}
\caption{The major latency composition.}\label{tab:pred_model_type}
\end{table}

\begin{table}[t]
	\centering
	\scriptsize
 	\begin{tabular}{|L{1cm}|L{3.8cm}|R{1cm}|R{1cm}|}
 	\hline
 	\textbf{Layer Type} & \textbf{Prediction Model} & \textbf{Latency Acc.} & \textbf{Energy Acc.}\\\hline
 	Conv & decision tree input: \textit{filter\_size}, linear regression input: \textit{batch $\cdot$ input\_width $\cdot$ input\_height $\cdot$ channel $\cdot$ kernel\_number $\div~stride^2$} & 0.993 & 0.973\\ \hline
 	Pooling & decision tree input: \textit{filter\_size}, linear regression input: \textit{batch $\cdot$ input\_width $\cdot$ input\_height $\cdot$ channel $\cdot$ kernel\_number $\div~stride^2$} & 0.784 & 0.772\\ \hline
 	Fully-connected & linear regression input: \textit{a\_width $\cdot$ a\_height $\cdot$ b\_width, a\_width $\cdot$ a\_height, b\_width $\cdot$ b\_height} & 0.945 & 0.922\\ \hline
 	Activation & decision tree input: \textit{activation function type}, linear regression input: \textit{input\_size} & 0.998 & 0.970\\ \hline
	\end{tabular}
\caption{Our latency \& energy prediction models for different kinds of DL layers and the prediction results.
We use Coefficient of Determination $R^2$ as the metric to evaluate the accuracy of our prediction models (best possible score is 1.0).
}
\label{tab:latency_prediction_result}
\end{table} 

\setlength{\textfloatsep}{10pt}
\begin{algorithm}[t]
\small
\KwIn{
$G$: pre-trained graph to be executed\\
$p(G)$: binary-partition function, returns a list of partitions $\langle$\textit{$G_{w}$, $G_{h}, dt$}$\rangle$, where \textit{dt} is the size of data to be transferred\\
\textit{f(G, $\mathcal{S}$), g(G, $\mathcal{S}$)}: pre-trained models for predicting the latency/energy of executing \textit{G} under device status \textit{s}\\
$\mathcal{S}_w$, $\mathcal{S}_h$: current device running status for wearable and handheld, including CPU frequency, CPU loads, etc\\
$\mathcal{B}$: current Bluetooth uplink bandwidth\\
\textit{PR, PT}: rx/tx power consumption over Bluetooth\\
\textit{PropT}: proper latency that the app is supposed to run on\\
$\mathcal{W}_w$, $\mathcal{W}_p$: weights of battery for wearable and handheld\\
}
\KwOut{Optimal partition choice}
$partitions \leftarrow p(G), L = E = \emptyset$\\
\ForEach{$\langle G_{w}, G_{h}, dt\rangle \in partitions$} {
  \uIf{\textit{streaming\_opt\_on}}{
    $l \leftarrow \max (f(G_{w}, \mathcal{S}_w), f(G_{h}, \mathcal{S}_p) + dt/\mathcal{B})$\\
  }
  \uElse{
    $l \leftarrow f(G_{w}, \mathcal{S}_w) + f(G_{h}, \mathcal{S}_p) + dt/\mathcal{B}$\\
  }
  $E_w \leftarrow g(G_{w}, \mathcal{S}_w) + dt * PT$\\
  $E_p \leftarrow g(G_{h}, \mathcal{S}_p) + dt * PR$\\
  $L.append(l), E.append(\mathcal{W}_w * E_w + \mathcal{W}_p * E_p)$\\
}
\uIf{PropT == 0 or $\min(L) > PropT$}{
  $opt\_index \leftarrow \argmin{i \in \{1...N\}} (L[i])$
}
\uElseIf{PropT == +$\infty$}{
  $opt\_index \leftarrow \argmin{i \in \{1...N\}} (E[i])$

}
\uElse{
  $\mathcal{R} \leftarrow$ list of index i that satisfies L[i] $\leqslant PropT$\\
  $opt\_index \leftarrow \argmin{i \in \mathcal{R}} (E[i])$
}
\Return{$partitions[opt\_index]$}\;
\caption{\framework Partition Algorithm.}
\label{alg:decision}
\end{algorithm}

\mysubsection{Making Offloading Decision}\label{sec:decision}
Utilizing the prediction models described above, \framework~dynamically selects the optimal partition point.
The decision making procedure involves two steps: finding a set of possible partitions for a given graph, and identifying the optimal one among them.

\begin{figure*}[t]
	\centering
	\includegraphics[width=0.75\textwidth]{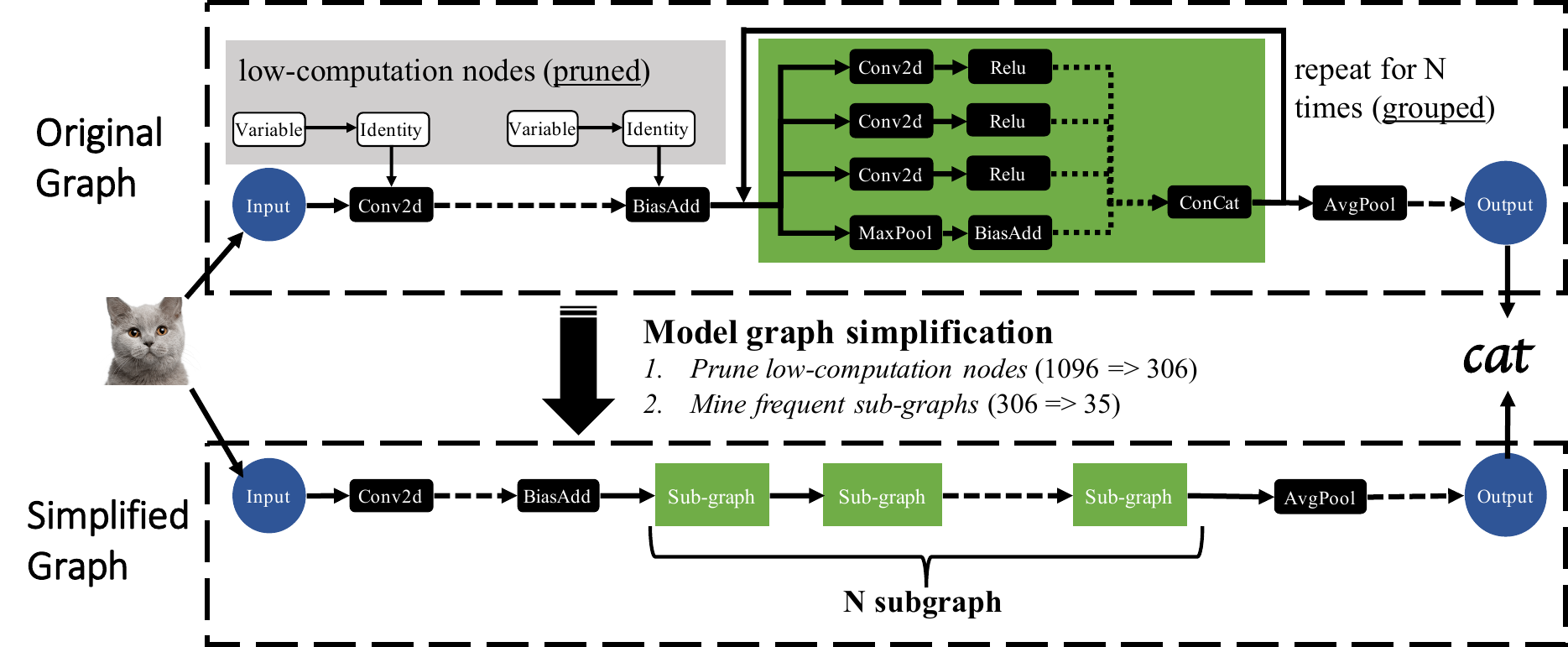}
	\caption{Example of how \framework simplifies \emph{GoogLeNet}.
  Each node presents a layer, while each edge presents the data flow among those layers. Dash lines indicate many more nodes are hidden to save space.
  \revise{\framework first prunes the model graph by keeping only the computation-intensive nodes (as listed in Table~\ref{tab:pred_model_type}), and then grouping the repeated subgraphs together. After these two steps, a complex directed acyclic graph often becomes a linear and much simpler graph.}
  }
	\label{fig:graph_simplification}
\end{figure*}

\textbf{Dynamic Partition.}
A DL model can be abstracted as a Directed Acyclic Graph (DAG) with the source (input) and the sink (output) nodes, where each node represents a layer and each edge represents the data flow among those layers.
A valid partition equals to a \emph{cut}~\cite{Cut} of the DAG and requires the source and the sink to be placed in different subsets.
Finding all \emph{cuts} of a given graph shall need the $\mathcal{O}(2^n)$ complexity where $n$ is the number of nodes. For a large DL model, e.g., the \emph{GoogLeNet} model with 1,096 nodes, such a complexity is prohibitive.
\revise{As pointed out previously by Kang \textit{et al.}~\cite{conf/asplos/KangHGRMMT17}, existing DL-partition approaches simply assume these graphs are linear.} Hence, each single edge represents a valid partition point.
However, we observe that such an assumption is not always true for many popular DL models (e.g., \emph{GoogLeNet}), as there can be branches and intersections in the graph.
This motivates us to design a heuristic-based algorithm that efficiently computes a set of ``representative'' cuts for a general graph of a DL model, as to be described below.

Figure~\ref{fig:graph_simplification} illustrates how our algorithm works.
First, \framework prunes all computationally light nodes, only keeping the computationally heavy nodes such as those shown in Table~\ref{tab:latency_prediction_result}.
After identifying these light nodes, \framework~removes them and connects their input nodes and output nodes.
Second, we observe that a DL (e.g., CNN and RNN) model often has repeated subgraph structures, which we call ``frequent subgraphs'', that frequently appear in the DAG.
\framework thus bundles each frequent subgraph into one virtual node without further splitting.
\mrev{To mine the frequent subgraphs, i.e., detecting the frequent patterns in a model and the nodes associated with those patterns,} the most straightforward way is to utilize the node \emph{namespace}: nodes under the same namespace are often in the same subgraph.
However, the idea of \emph{namespace} is not supported in all DL frameworks;
more importantly, setting the namespaces is rather subjective and optional, and requires developers' additional support.
We utilize GRAMI~\cite{DBLP:journals/pvldb/ElseidyASK14}, a fast and versatile algorithm that automatically mines frequent subgraphs.
After the graph is simplified, there will be much fewer nodes (e.g., 1,096 to 35 for \emph{GoogLeNet}).
Additionally, the graph exhibits a mostly linear structure.
This allows us to apply a brute-force approach to identifying all \emph{cuts}. 
In addition, this simplification results can be cached for every single DL model and reused. We empirically observed that our heuristic-based partition identification approach is effective and robust.

\textbf{Optimal Partition Selection.}
The algorithm for determining an  optimal partition is demonstrated in Algorithm~\ref{alg:decision}.
Taking as input possible partitions generated in the previous step,
\framework~analyzes the partitioned subgraphs on the wearable and the handheld, and uses the prediction models (Section~\ref{sec:prediction_model}) to estimate the corresponding latency and energy consumption (line 2$\sim$10).
Note that the overall energy consumption metric is a weighted mean from the energy consumed on both the wearable and handheld.
Our algorithm provides a general framework for diverse usage scenarios.
If the DL app integrated with \framework~is latency-sensitive specified by developers, we select the partition with the smallest latency (line 11$\sim$12).
In contrast, if the app is latency-insensitive, then we select the partition with the lowest energy consumption (line 13$\sim$14).
In a more general case, the developer is able to quantitatively specify the latency requirement. We then select the most energy-efficient partition satisfying this requirement (line 15$\sim$17).

The models and parameters in Algorithm~\ref{alg:decision} are obtained from various sources and can be classified into four types:
(1) offline-training models, including the latency and energy prediction models \textit{(f, g)}, as well as the power model of Bluetooth data transfer \textit{(PR, PT)},
(2) runtime-profiling parameters gathered by the \textit{System Profiler} module (Section~\ref{sec:overall_architecture}), including the handheld status $\mathcal{(S)}$ and Bluetooth bandwidth \textit{(B)},
(3) application-specified parameters, including the expected end-to-end latency of DL inference \textit{(PropT)},
(4) configurable trade-off parameters, including energy consumption weights on wearable and handheld \textit{($\mathcal{W}_w$, $\mathcal{W}_p$)}.

\mysubsection{Optimizing Streaming Data Processing}\label{sec:streaming}
DL tasks such as video stream analysis for augmented reality and speech recognition will become common on wearable devices. In these tasks, the input consists of a stream of data such as video frames and audio snippets that are continuously fed into the same model. Here we use ``frame'' to denote an input unit for a DL model, e.g., an image or an audio snippet. Compared to non-streaming data, streaming data cares more about the overall throughput, i.e., how many frames can be processed per time unit, rather than the latency for every single frame.
For the non-streaming input, the data dependency between two partitioned sub-models makes pipelined or parallel processing impossible: when the wearable is processing the first part of the model, the handheld has to wait for its output that serves as the input to the second part of the model to be executed on the handheld.
For streamed input, however, \framework employs \emph{pipelined processing} on wearable and handheld. Specifically, when the $n$-th frame finishes computing on the wearable and being sent to the handheld, the wearable can immediately start processing the $(n+1)$-th frame, and so on.

Pipelining helps fully utilize the computation resources on both devices and thus effectively improves the overall throughput.
To integrate the pipelining support into our partition-decision algorithm, we revise the end-to-end latency calculation in Algorithm 3.1 as the maximum of the wearable computation delay and the handheld computation delay along with the data transmission delay (Line 4).
In other words, due to pipelining, the amortized end-to-end latency is determined by the processing delay on either device, whichever is longer.

\subsection{Provided Developer APIs}\label{sec:developer_APIs}

\lstset{frame=tb,
  language=Java,
  aboveskip=3mm,
  belowskip=3mm,
  showstringspaces=false,
  columns=flexible,
  basicstyle={\small\ttfamily},
  numbers=left,
  numberstyle=\tiny\color{gray},
  keywordstyle=\color{blue},
  commentstyle=\color{dkgreen},
  stringstyle=\color{mauve},
  breaklines=true,
  breakatwhitespace=true,
  tabsize=3,
  xleftmargin=1cm
}

\framework exposes a set of easy-to-use APIs for developers for running the model inference, as listed in the code snippet in List~\ref{list:javaSample}. The high-level design principle of such APIs is to minimize the developers' additional overhead including learning curve and programming efforts.
Therefore, low-level details of whether/when/how to offload should be completely transparent to developers.
As a result, the exposed interfaces are almost the same as a conventional DL library such as TensorFlow.
The only new knob \framework provides is a hint function for specifying the latency requirement (Line 3 in List~\ref{list:javaSample}), which helps \framework make offloading decisions.

\begin{lstlisting}[caption={A code sample of using \framework}, label={list:javaSample}]
DeepWearInference infer =
  new DeepWearInference("/path/to/model");
infer.set_expected_latency(100); // 100ms
infer.feed(input_name, input_data);
infer.run();
float[] result = infer.fetch(output_name);
\end{lstlisting}

As exemplified in the code snippet~\ref{list:javaSample},
using the APIs provided by \framework is quite similar to using the standard Java APIs~\cite{TFAPIs} provided by TensorFlow.  
It typically consists of four steps: loading pre-trained model, feeding the input, executing the graph, and finally fetching the output.
Unlike traditional general-purpose offloading frameworks, \framework doesn't require any manual annotation to specify what can be offloaded. In contrast, \framework hides the offloading details from the perspective of developers. 

\mysection{Implementation of \framework}\label{sec:implementation}
We have implemented \framework on commodity smartphone and smartwatches running off-the-shelf Android and Android Wear OS respectively.
Our prototyping efforts consist of around 3,200 lines of code written in Java, excluding the scripts for constructing and analyzing prediction models.
Developers can easily integrate \framework into their apps by importing the \framework library on both the wearable side and the handheld side.
In the one-time initialization phase when the app is being installed, \framework will also locate other necessary components such as the DL models (stored at both the wearable and the handheld) and the latency/energy prediction models (stored at the wearable).
The handheld-side library also provides a console allowing users to configure offloading policies as described in Section~\ref{sec:decision}).

Currently, \framework~employs the popular TensorFlow~\cite{TensorFlow} as our DL algorithms driver (Figure~\ref{fig:architecture}).
Other popular frameworks such as Caffe2~\cite{caffe2} and PyTorch~\cite{pytorch} can also be easily integrated into \framework with very small adaptation.
To realize the System Profiler, \framework obtains the processor status via the \emph{sysfs} interface.
More specifically, the CPU information can be obtained from~\texttt{/sys/devices/system/cpu/} on both the smartphone and the smartwatch.
For GPU on smartphones, the hardware driver exposes the information such as the total running time and busy time. On the Nexus 6 model, such information can be obtained from \texttt{/sys/class/kgsl/kgsl-3d0/}.
The data communication between wearable and handheld is realized by the standard Android Wearable Data Layer API~\cite{DataLayer}. 
Specifically, we use the \textit{Message API}~\cite{MessageAPI} for the message exchange in the control channel, and use \textit{DataItem \& Asset APIs} for transferring computation results and intermediate data (when the DL model is partitioned across the two devices).
The Bluetooth bandwidth profiling is performed either passively (by measuring the offloaded data transfer) or actively (by sending lightweight probing packets). The active probing is triggered
periodically (every 1 minute by default) as well as by Bluetooth signal strength changes, in the absence of offloading transfers.
We are currently working on adding Direct WiFi support for offloading.

\begin{table*}[!t]
\centering
\scriptsize
\begin{tabular}{|l|l|l|l|l|l|l|l|l|l|}
\hline
\multirow{2}{*}{\textbf{Wearable}} & \multirow{2}{*}{\textbf{Handheld}} & \multicolumn{8}{c|}{\textbf{Models}}\\\cline{3-10}
& & \emph{MNIST} & \emph{GoogLeNet} & \emph{MobileNet} & \emph{WaveNet} & \emph{LSTM-HAR} & \emph{DeepSense} & \emph{TextRNN} & \emph{DeepEar}\\\hline
\multirow{3}{*}{\textbf{LG Urbane}} & \textbf{CPU-interactive} & input & input & Squeeze & input & input & input & BiasAdd & output\\\cline{2-10}
& \textbf{CPU-powersave} & add\_3 & \cellcolor{red}AvgPool\_0a & Squeeze & input & input & input & BiasAdd & output\\\cline{2-10}
& \textbf{GPU} & input & input & Squeeze & input & input & input & BiasAdd & output\\\cline{1-10}
\multirow{3}{*}{\textbf{Galaxy S2}} & \textbf{CPU-interactive} & add\_3 & AvgPool\_0a & Squeeze & input & input & input & BiasAdd & output\\\cline{2-10}
& \textbf{CPU-powersave} & add\_3 & Squeeze & Squeeze & logit/out & input & input & BiasAdd & output\\\cline{2-10}
& \textbf{GPU} & add\_3 & Squeeze & Squeeze & input & input & input & BiasAdd & output\\\hline
\end{tabular}
\caption{\framework~partition point selections under different devices and models ($PropT = 0$). Red blocks indicate \framework~fails to make the optimal partition choice and white block means the optimal partition point is picked.
}\label{tab:selection_accuracy}
\end{table*}

\begin{figure*}[t]
	\centering
	\includegraphics[width=0.98\textwidth]{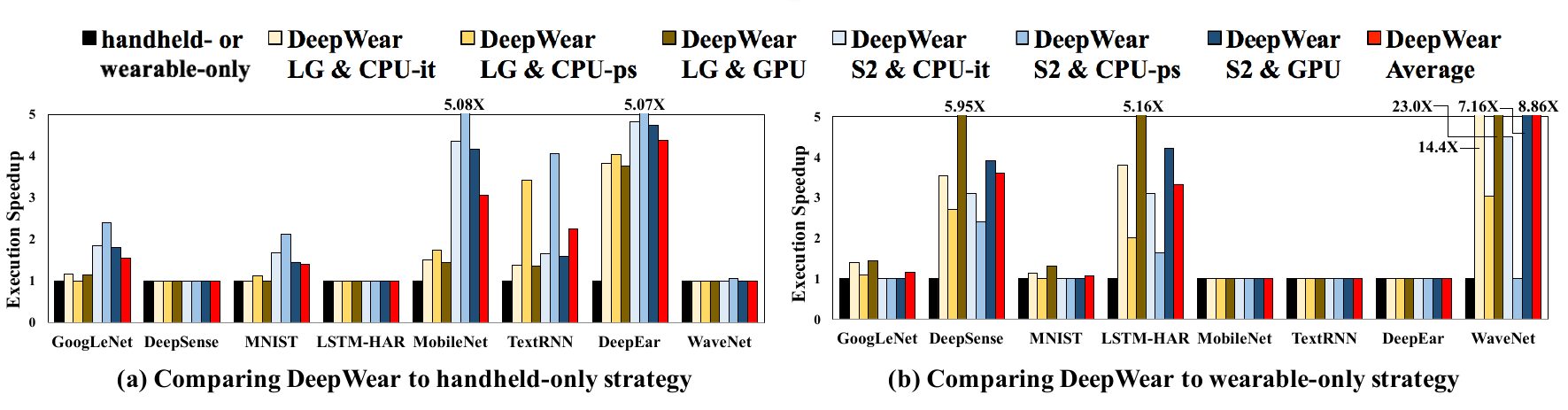}
	\caption{Normalized execution speedup of \framework to two naive strategies: handheld-only and wearable-only.
	We present the results under 6 configurations for wearables (LG Urbane and Galaxy S2) and handhelds (CPU-interactive, CPU-powersave, and GPU). Note that numbers shown represent the relative speedup, with the handheld/wearable-only being the comparison baseline.
	We use configuration $PropT = 0$, so that \framework~will chase for the smallest end-to-end latency.
	Results show that \framework~can improve the latency by 1.95X and 2.62X on average compared to wearable-only and handheld-only, respectively. Additionally, the improvement can be up to 5.08X and 23.0X in some cases.
	}
\label{fig:latency_improvements}
\end{figure*}

\begin{figure*}[t]
	\centering
	\includegraphics[width=0.98\textwidth]{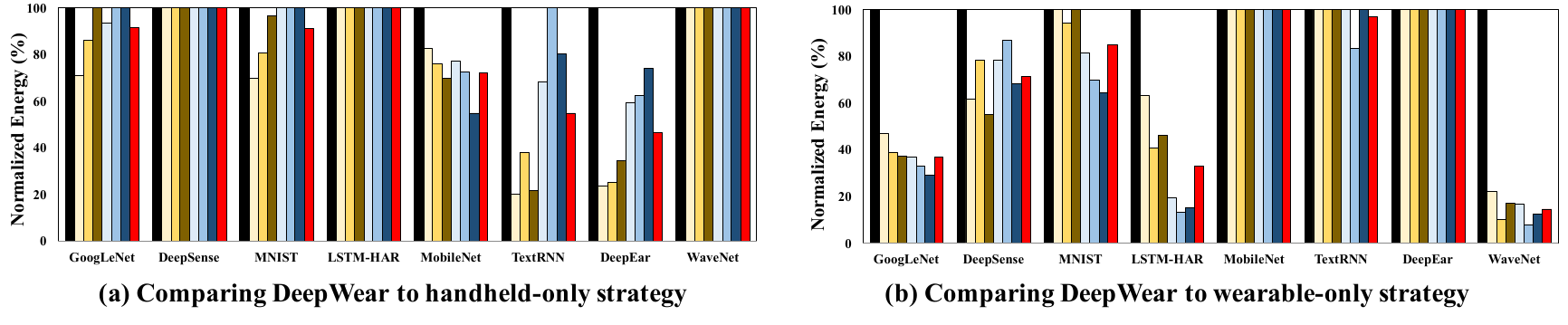}
	\caption{Normalized energy consumption of \framework to two naive strategies: handheld-only and wearable-only.
	We present the results under 6 running scenarios about wearables (LG Urbane and Galaxy S2) and handhelds (CPU-interactive, CPU-powersave, and GPU).
	We use configuration $PropT = +\infty, \mathcal{W}_w = \mathcal{W}_p = 0.5$, so that \framework~will chase for the smallest energy consumption.
	Results show that \framework~can save energy by 18.0\% and 32.7\% on average compared to the handheld-only and the wearable-only, respectively. Additionally, the improvement can be up to 53.5\% and 85.5\% in some cases. \mrev{Note that the handheld energy consumption is calibrated using the method used for Figure~\ref{fig:inception_partition}, in order to take into consideration the phone and wearable's heterogeneous battery capacities.}}
\label{fig:energy_improvements}
\end{figure*}
\mysection{Evaluation}\label{sec:eval}
We now comprehensively evaluate \framework~using the aforementioned 8 popular DL models under different device configurations.
The experimental setup is the same as that used in Section~\ref{sec:back}.
Each experiment is repeated for 20 times to make the results statistically meaningful.

\mysubsection{Partition Selection Accuracy}\label{sec:eval_accuracy}
Table~\ref{tab:selection_accuracy} shows the partition points selected by \framework~under different devices and DL models.
Each cell represents the DL layer name at which \framework performs the partition, indicating that the output data of this layer shall be offloaded to the handheld.
The red block indicates that \framework~fails to make the optimal partition choice.
Here, an ``optimal'' partition choice means that it outperforms all other partition choices for the specified goal, e.g., end-to-end latency when \textit{PropT} equals to 0 in our case.
\mrev{We obtain the optimal partition choice by exhaustively testing each possible partition point.}
In summary, \framework~is able to select the best partition point for 47 out of 48 cases we tested (97.9\%).
The mis-predictions occur because of two reasons.
First, our prediction models used in \framework consider only a subset of layer types as explained in Section~\ref{sec:prediction_model}.
Second, those prediction models themselves cannot perfectly predict the delay or energy.
Also note that for all 3 suboptimal partition points in Table~\ref{tab:selection_accuracy}, their delay and energy consumption are actually very close to those of the optimal partitions.

\mysubsection{Latency and Energy Improvements}\label{sec:eval_improvements}
To demonstrate how \framework~can help improve the end-to-end latency and energy consumption, we test it under two extreme cases: optimizing for latency ($PropT = 0$) and optimizing for energy ($PropT = +\infty$).
We present the results under 6 running scenarios about wearables (LG Urbane and Galaxy S2) and handhelds (CPU-interactive, CPU-powersave, and GPU).
We compare the performance of \framework~with two baseline strategies: handheld-only (offloading all tasks to the handheld) and wearable-only (executing the entire model on the wearable without performing offloading).

\textbf{Speedup.}
Figure~\ref{fig:latency_improvements} shows \framework's execution speedup (normalized) over the baseline strategies across 8 DL models and varied device specifications \& status.
Bars in different colors represent different hardware configurations.
LG and S2 are abbreviated for Urbane LG and Galaxy S2.
CPU-it, CPU-ps, and GPU refer to utilizing Nexus 6 under CPU-interactive, CPU-powersave, and GPU on the handheld side, respectively.
The black bar represents the latency of handheld- or wearable-only approaches, and is used as a baseline to normalize other approaches (normalized to 1 itself).
The red bar is the average speedup for each model.

Figure~\ref{fig:latency_improvements}(a) shows that compared to the handheld-only strategy, \framework can help reduce the latency for 6 out of 8 models, with an average improvement ranging from 1.01X (\emph{WaveNet}) to 4.37X (\emph{DeepEar}).
Similarly, Figure~\ref{fig:latency_improvements}(b) shows that compared to the wearable-only strategy, \framework~reduces the latency of running 5 out of 8 models with an average improvement ranging from 1.07X (\emph{MNIST}) to 8.86X (\emph{WaveNet}).
For cases such as running \emph{WaveNet} on LG Urbane with Nexus GPU 6 available, \framework~can even speed up the processing for more than 20 times (23.0X) compared to the wearable-only strategy.
Overall, \framework~can improve the latency by 2.62X and 1.95X on average compared to wearable-only and handheld-only, respectively, across all 8 models.

Our another observation is that different models can exhibit quite diverse results.
We find that the execution speedup achieved by \framework depends on two factors related to the model structure: computation workloads and data size.
A model graph with small computation workloads (\emph{DeepEar}, \emph{TextRNN}) or with a large input data size (image-processing applications such as \emph{GoogLeNet} and \emph{MobileNet}) are unlikely to benefit from the offloading since the performance bottleneck often resides in the data transmission rather than the local processing.
Hence, in these cases, \framework can have significant improvements over the handheld-only approach, but less improvement over the wearable-only approach.
In contrast, when running DL models that require lots of computations on a relatively small size of data, \framework exhibits more improvements compared to the wearable-only approach rather than the handheld-only approach.

\textbf{Energy saving.}
Similarly, Figure~\ref{fig:energy_improvements}(a) shows that compared to handheld-only, \framework~can help reduce the energy consumption for 5 out of 8 models, with an average improvement (the red bar) ranging from 8.3\% (\emph{GoogLeNet}) to 53.5\% (\emph{DeepEar}).
Similarly, Figure~\ref{fig:energy_improvements}(b) illustrates that compared to wearable-only, \framework~lowers the energy consumption for 6 out of 8 models, with an average improvement ranging from 3.8\% (\emph{TextRNN}) to 85.5\% (\emph{WaveNet}).
Overall, \framework~can on average save the energy by 18.0\% and 32.7\% compared to the handheld-only and the wearable-only approach, respectively.


\mysubsection{Local Offloading vs. Cloud Offloading}
We also compare \framework's local offloading approach to offloading to the remote cloud.
We use a server equipped with Tesla K80 GPU, 2.3GHz Intel Xeon CPU, and 60GB memory to play as the remote cloud.
We carry out the experiments under two WiFi conditions: poor ($\sim$ 100kbps) and good ($\sim$ 5mbps).
\footnote{We notice wearable's WiFi connectivity is oftentimes slower than phone due to wearable's form factor (smaller antenna).}
The results are all normalized by the wearable-only performance (no offloading).
Note that for cloud offloading, we ignore the cloud server's energy consumption.

\begin{figure}[t]
\subfloat[Normalized execution speedup.]{\includegraphics[width=0.48\textwidth]{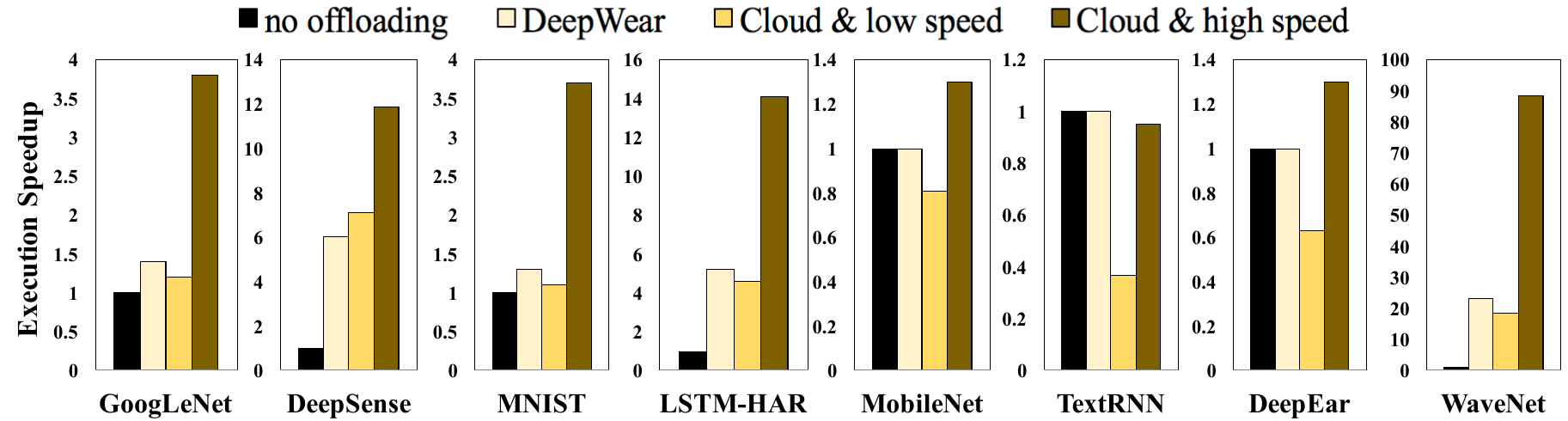}}\\
\subfloat[Normalized energy consumption. \revise{For \framework, we measure the \revise{total energy consumption} of both wearable and handheld. For the cloud offloading, we measure only the energy consumption on the wearable.}]{\includegraphics[width=0.48\textwidth]{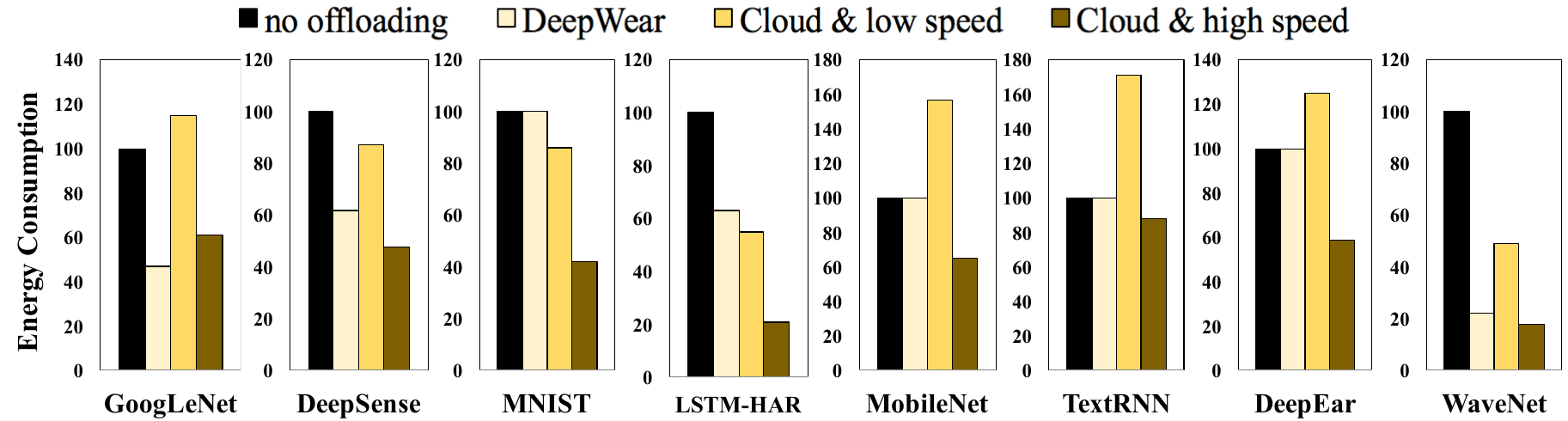}}
\caption{Compare \framework to cloud offloading}
\label{fig:cloud}
\end{figure}

For latency improvements, as shown in Figure~\ref{fig:cloud}(a), cloud offloading outperforms both local execution and \framework under good network condition.
However, when the network condition becomes poor, cloud offloading is comparable and sometimes performance-wise worse than \framework.
Regarding the energy consumption, as shown in Figure~\ref{fig:cloud}(b), \framework can even outperform cloud offloading under good network condition (\emph{WaveNet}).
The reason is that the Internet access over WiFi is more energy-consuming than accessing the handheld over local radio.
Note that under cellular network (e.g., LTE) the energy consumption can be even more than WiFi, thus \framework is expected to exhibit more improvements.
Finally, recall that compared to cloud offloading, \framework offers other benefits such as ubiquitousness and better privacy as described in Section~\ref{sec:intro}.

It's worth mentioning that even though the cloud offloading may perform better than \framework under many circumstances, offloading user data to cloud still suffers from privacy concerns, since these data such as images, sensor output, and audio used for these DL models often contain sensitive personal information. Since \framework instead performs local offloading, it reduces such privacy concerns to the minimum.

\begin{figure}[t]
\centering
\scriptsize
\subfloat[\emph{TextRNN} model]{\includegraphics[width=0.25\textwidth]{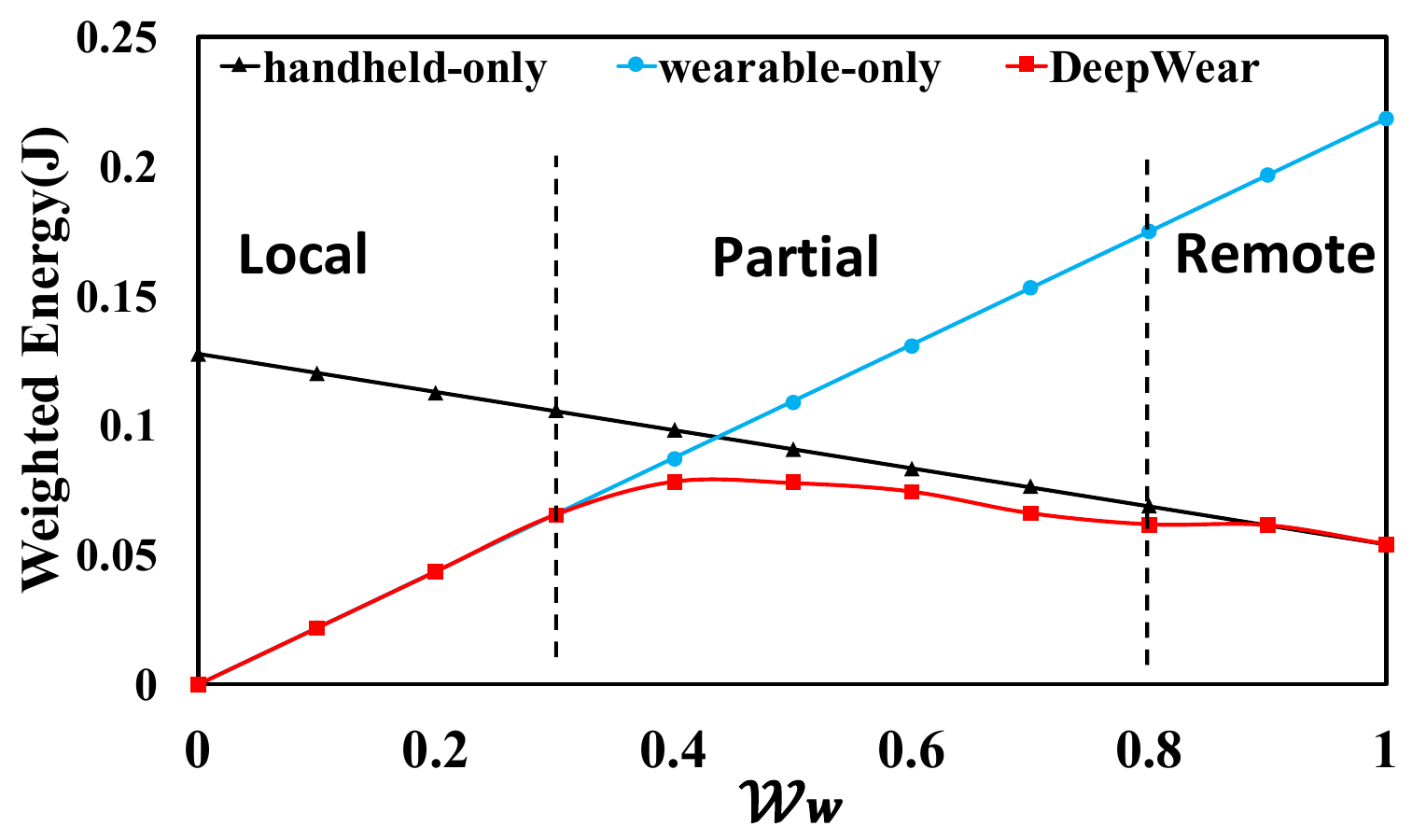}}
\subfloat[\emph{MobileNet} model]{\includegraphics[width=0.25\textwidth]{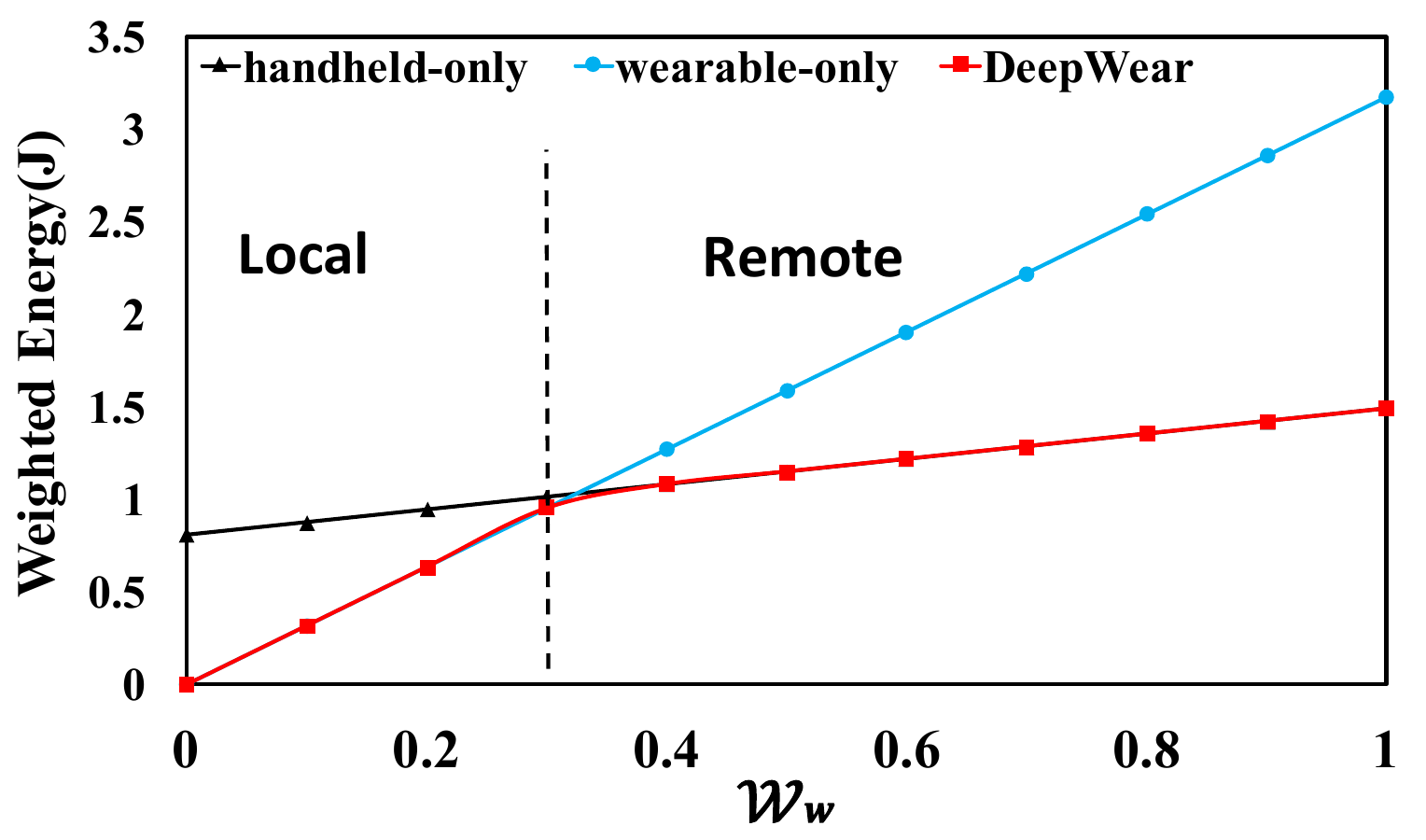}}
\caption{
\revise{Weighted energy consumption for different $\mathcal{W}_w$ and $\mathcal{W}_p = 1 - \mathcal{W}_w$.
$\mathcal{W}_w$ and $\mathcal{W}_p$ are the energy weight of the wearable and the handheld, respectively.
The Y-axis represents the weighted sum of the energy consumption of both devices as $\mathcal{W}_w \cdot E_w + \mathcal{W}_p \cdot E_p$. We use Galaxy S2 and Nexus 6 CPU-powersave to carry out this experiment.}
}
\label{fig:weight}
\end{figure}

\begin{figure}[t]
\centering
\scriptsize
\subfloat[\emph{MNIST} model]{\includegraphics[width=0.25\textwidth]{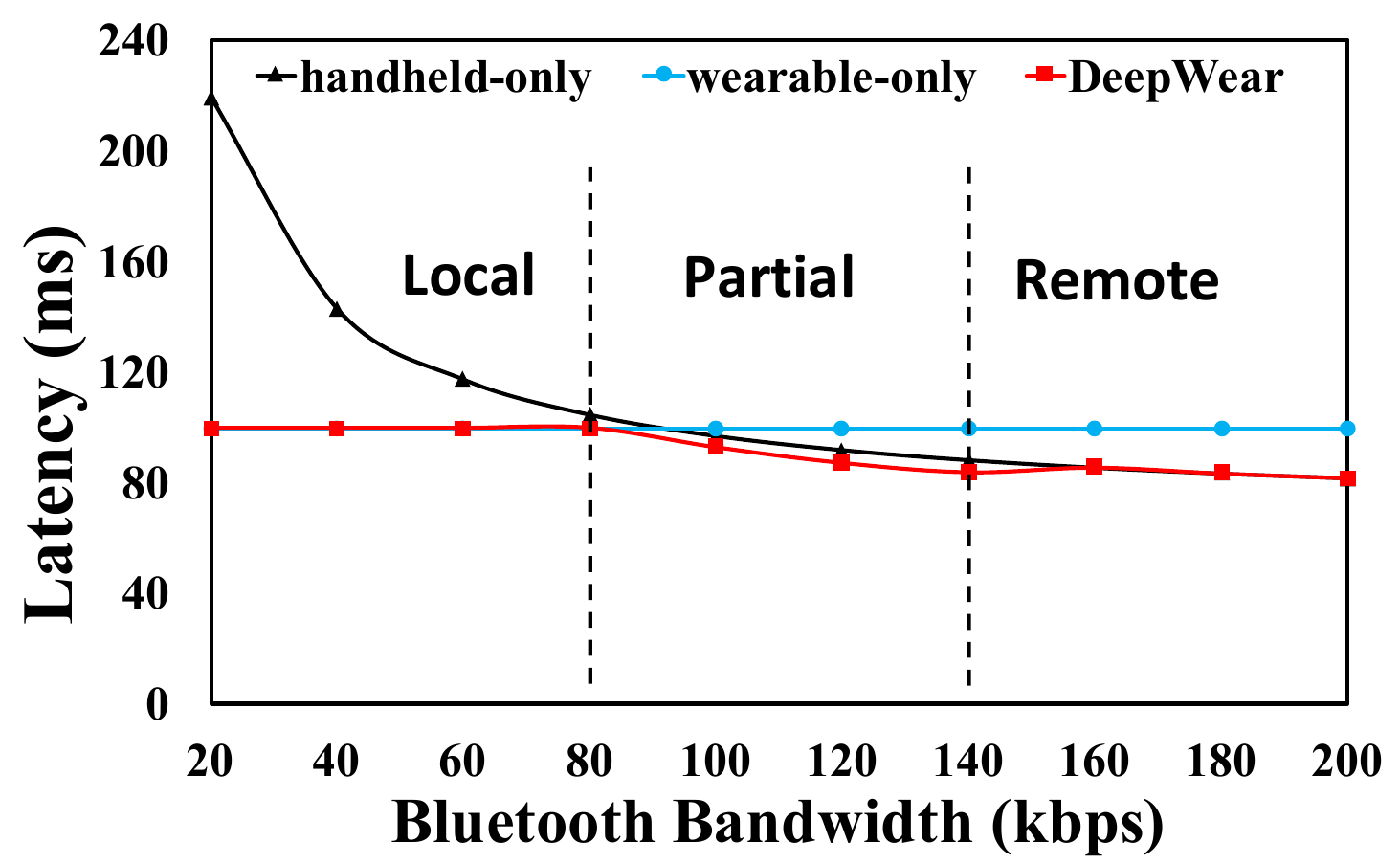}}
\subfloat[\emph{GoogLeNet} model]{\includegraphics[width=0.25\textwidth]{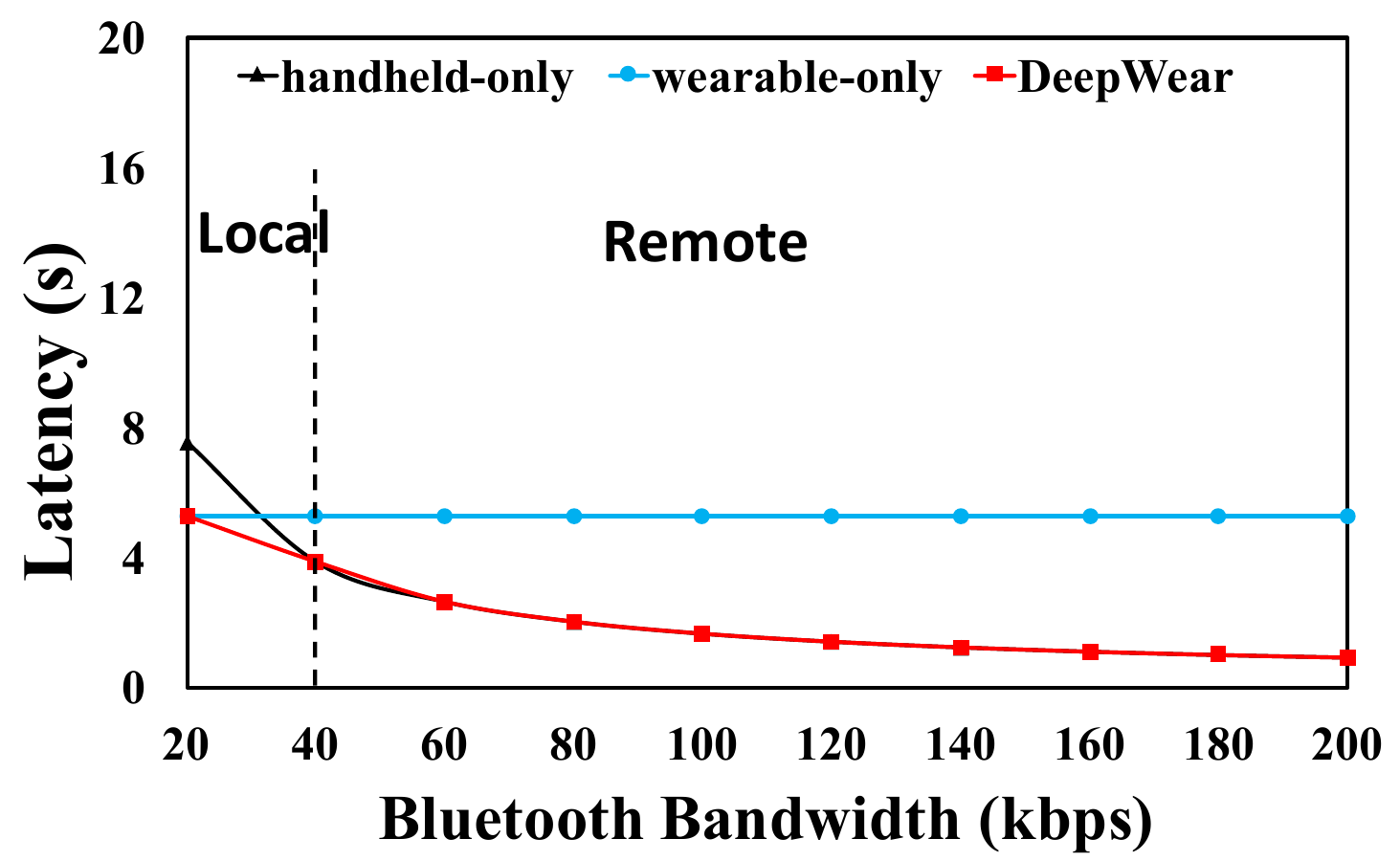}}
\caption{End-to-end latency across different $\mathcal{B}$.
\revise{We use Urbane LG and Nexus 6 CPU-interactive to carry out this experiment.}
}
\label{fig:bt}
\end{figure}

\begin{figure}[t]
\centering
\scriptsize
\subfloat[\emph{MNIST} model]{\includegraphics[width=0.25\textwidth]{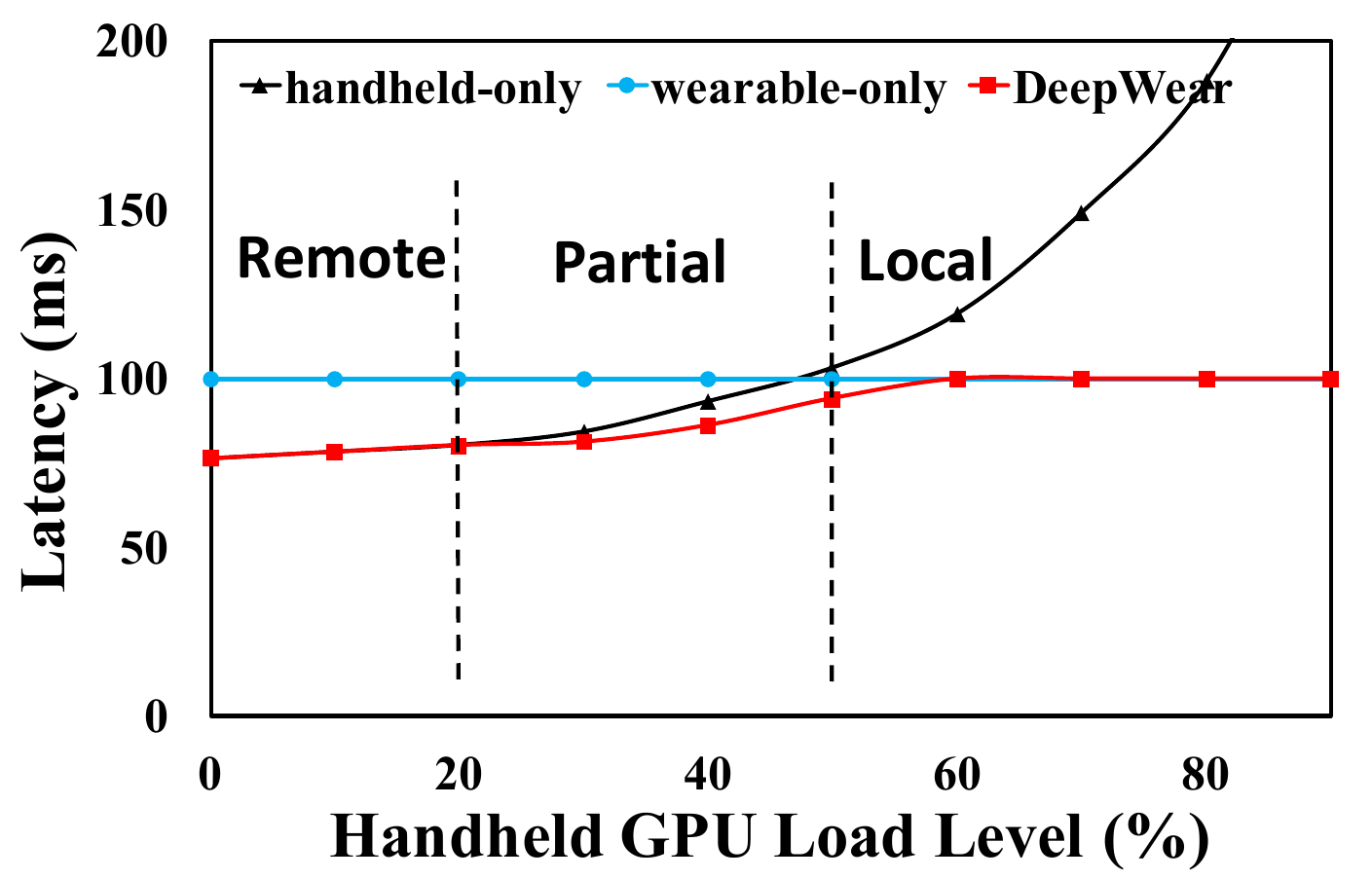}}
\subfloat[\emph{LSTM-HAR} model]{\includegraphics[width=0.25\textwidth]{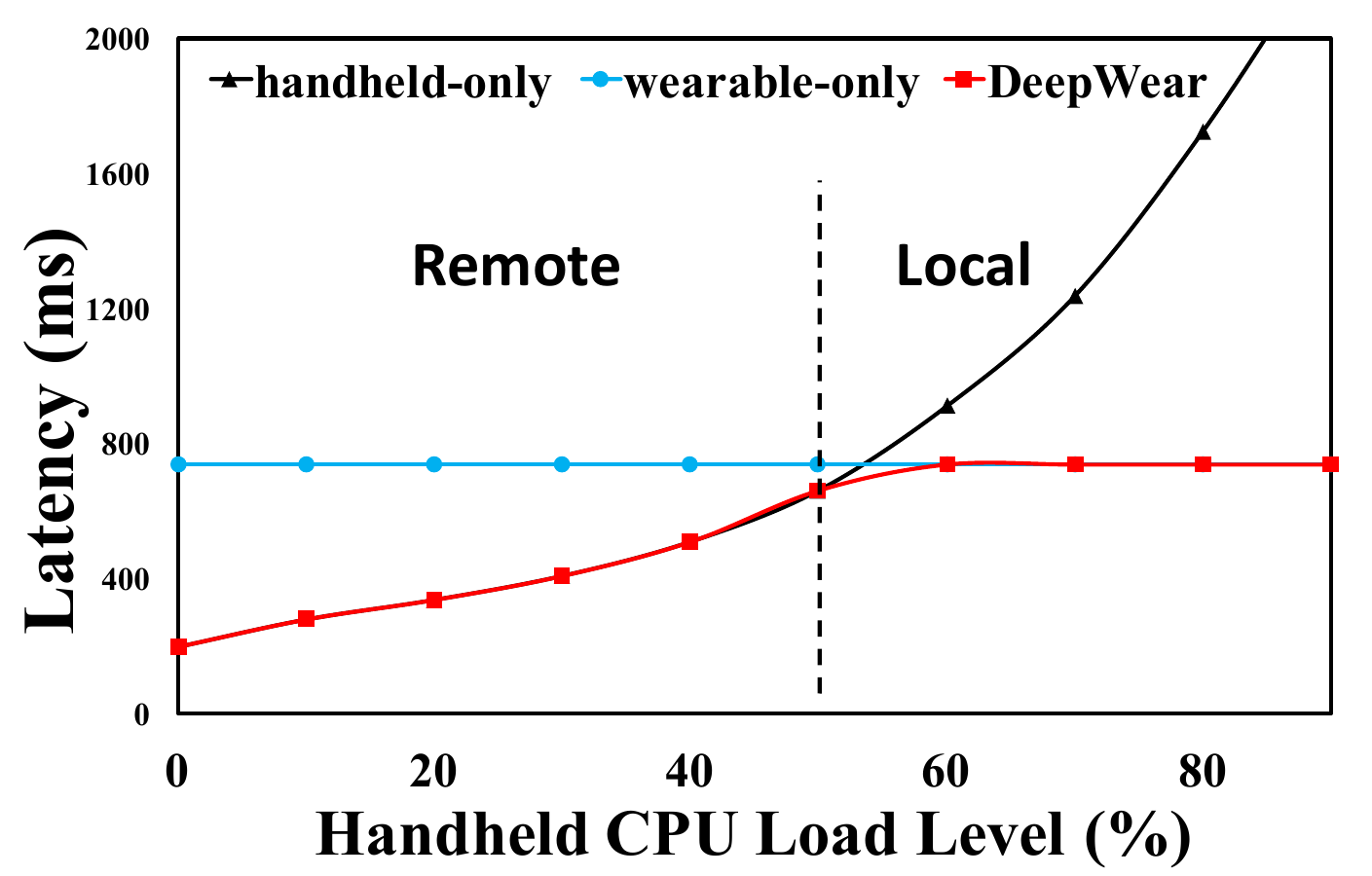}}
\caption{End-to-end latency across different $\mathcal{S}$.
\revise{We use the Urbane LG and the Nexus 6 (GPU and CPU) to carry out this experiment.}
}
\label{fig:load}
\end{figure}

\mysubsection{Adaptive to Environment Dynamics}\label{sec:eval_adaptiveness}
In this section, we evaluate \framework's adaptiveness to diverse factors that may vary in real-world environments: the device battery level ($\mathcal{W}_w, \mathcal{W}_p$), the Bluetooth bandwidth (\textit{B}), and the processor load level ($\mathcal{S}_p$).
Our experimental results show that \framework can effectively adapt to the dynamics caused by these external factors.

\textbf{Battery level.}
As mentioned in Section~\ref{sec:decision}, \framework's offloading decision should consider the battery level of both the wearable and the handheld, in order to better balance their battery life.
This is achieved by tuning the parameters $\mathcal{W}_w$ and $\mathcal{W}_p$.
We exemplify a possible policy as follows.
When the handheld is being charged, we focus on saving the energy for wearable (i.e., $\mathcal{W}_w = 1, \mathcal{W}_p = 0$),
whereas when the handheld's battery is running out, we should more aggressively use the wearable's battery (e.g., by setting $\mathcal{W}_w = 0.2$ and $\mathcal{W}_p = 0.8$).

We test \framework's robustness against the varying values of $\mathcal{W}_w$ and  $\mathcal{W}_p$ (set to $1-\mathcal{W}_w$).
As shown in Figure~\ref{fig:weight},
the partition decision of \framework~keeps changing according to the configuration of energy weight.
As a result, \framework~always consumes no more energy than either the wearable-only or the handheld-only strategy.
Taking \emph{TextRNN} as an example, when $\mathcal{W}_w$ is low (0 $\sim$ 0.3), \framework~chooses to run the model locally as the wearable energy is relatively ``cheap''.
When $\mathcal{W}_w$ becomes higher (0.3 $\sim$ 0.8), the model is partitioned and executed on both sides.
During this stage, \framework~outperforms both wearable-only and handheld-only strategies.
When $\mathcal{W}_w$ is high, \framework~offloads all workloads to the handheld to save the energy of wearable.
The results of \emph{MobileNet}, another example shown in Figure~\ref{fig:weight}(b), are similar to \emph{TextRNN}, except that for \emph{MobileNet} there is no partial offloading stage.
Such a difference stems from the different internal structure of \emph{MobileNet}.

\textbf{Bluetooth bandwidth.}
The Bluetooth bandwidth between the wearable and the handheld can change dynamically according to their distance.
\framework~profiles and takes into account this bandwidth online for the partition decision.
Figure~\ref{fig:bt} shows how \framework reacts to the changing bandwidth in consideration of end-to-end latency.
As observed from both Figure~\ref{fig:bt}(a) (the \emph{MNIST} model) and Figure~\ref{fig:bt}(b) (the \emph{GoogLeNet} model), \framework~tends to execute the whole DL model locally when the bandwidth is low; 
when the bandwidth is high, \framework~performs offloading more aggressively.
Additionally, \framework~also chooses to partially offload the workload. For example, when running \emph{MNIST} with a bandwidth of 100kbps to 140kbps, partial offloading leads to better performance than both the wearable-only and the handheld-only strategies.

\textbf{Handheld processor load level.}
We then evaluate \framework's robustness against varying load level of the handheld processors (CPU and GPU).
We use a script~\cite{CPUloadgen} to generate CPU workloads, and use another application~\cite{huang2017shuffledog} to generate GPU workloads by introducing  background graphics rendering.
As shown in Figure~\ref{fig:load}, when the processor load is low, \framework always offloads the DL tasks to handheld to make use of the under-utilized processing power.
In this stage, the performance of \framework~is similar to handheld-only, and has significant latency reduction compared to wearable-only
(e.g., more than half a second for \emph{LSTM-HAR} model shown in Figure~\ref{fig:load}(b)).
When the handheld processor's load increases, \framework~chooses to execute workloads locally on the wearable device, as doing so outperforms the handheld-only approach.
For example, when running \emph{MNIST} with the handheld GPU load of 80\%, \framework~can reduce almost 50\% of the latency compared to the handheld-only strategy (188.2ms vs. 99.8ms).

\begin{table*}[]
\centering
\scriptsize
\begin{tabular}{|l|l|l|l|l|L{1cm}|l|l|l|l|l|}
\hline
\multicolumn{1}{|c|}{\multirow{2}{*}{\textbf{Model}}} & \multicolumn{1}{c|}{\multirow{2}{*}{\textbf{\textit{PropT}}}} & \multicolumn{3}{c|}{\textbf{handheld-only}} & \multicolumn{3}{c|}{\textbf{wearable-only}} & \multicolumn{3}{c|}{\textbf{DeepWear}} \\ \cline{3-11}
\multicolumn{1}{|c|}{} & \multicolumn{1}{c|}{} & Selection & Latency(ms) & Energy(mJ) & Selection & Latency(ms) & Energy(mJ) & Selection & Latency(ms) & Energy(mJ) \\ \hline
\multirow{3}{*}{\emph{TextRNN}} & 200ms & \multirow{3}{*}{input} & \multirow{3}{*}{239.60} & \multirow{3}{*}{181.79} & \multirow{3}{*}{BiasAdd} & \multirow{3}{*}{58.90} & \multirow{3}{*}{218.28} & BiasAdd & 58.90 & 218.28 \\ \cline{2-2} \cline{9-11}
 & 250ms &  &  &  &  &  &  & \begin{tabular}[c]{@{}l@{}}cell/mul\end{tabular} & 238.78 & 194.32 \\ \cline{2-2} \cline{9-11}
 & 300ms &  &  &  &  &  &  & \begin{tabular}[c]{@{}l@{}}Sigmoid\end{tabular} & 256.90 & 179.45 \\ \hline
 \emph{GoogLeNet} & 2s$\sim$3s & input & 7,306.21 & 9,361.00 & Squeeze & 2,058.50 & 7,616.64 & Squeeze & 2,058.50 & 7,616.64 \\\hline
\emph{LSTM-HAR} & 1s$\sim$2s & input & 207.02 & 317.27 & output & 733.53 & 1,207.26 & input & 207.02 & 317.27 \\\hline
\end{tabular}
\caption{End-to-end latency and energy consumption (\revise{of both the wearable and handheld}) of \framework~across varied developers-specified latency (\textit{PropT}). We use Galaxy S2 and Nexus 6 CPU-powersave to carry out this experiment. We set $\mathcal{W}_w$ and $\mathcal{W}_p$ as 0.5 equally.}
\label{tab:propt}
\end{table*}

\mysubsection{Latency Awareness}\label{sec:eval_awareness}
We also evaluate how the developer-specified latency (\textit{PropT}) affects \framework's decision on offloading.
The results are shown in Table~\ref{tab:propt}.
Overall, for 7 out of 9 configurations, \framework~can satisfy the latency requirement, while the handheld-only and the wearable-only have only 4 and 6, respectively.
The only case where \framework~is unable to provide the desired latency improvement, i.e., \textit{PropT} = 2.0s for \emph{GoogLeNet}, is unavoidable since even the lowest possible latency is higher than \textit{PropT}.
In those cases, \framework~chooses to minimize the end-to-end latency and ignore the energy consumption.
In summary, in all cases, \framework yields satisfactory results.

Another key observation from Table~\ref{tab:propt} is that \framework can adaptively adjust its decisions based on applications' requirements -- a desirable feature in practice.
Taking \emph{TextRNN} as an example. When \textit{PropT} is low, \framework keeps all workloads on the local wearable device to satisfy (58.9ms) the latency requirement (200ms).
This is the same as what the wearable-only strategy does but the handheld-only strategy fails to achieve.
When \textit{PropT} becomes higher (300ms), \framework~chooses different partition points in order to consume less energy than the wearable-only strategy, while keeping a relatively low end-to-end latency.
Instead, the wearable-only strategy consumes 21.6\% more energy than \framework.

\mysubsection{Handling Streaming Data}\label{sec:eval_streaming}

\begin{figure}[t]
	\centering
	\includegraphics[width=0.48\textwidth]{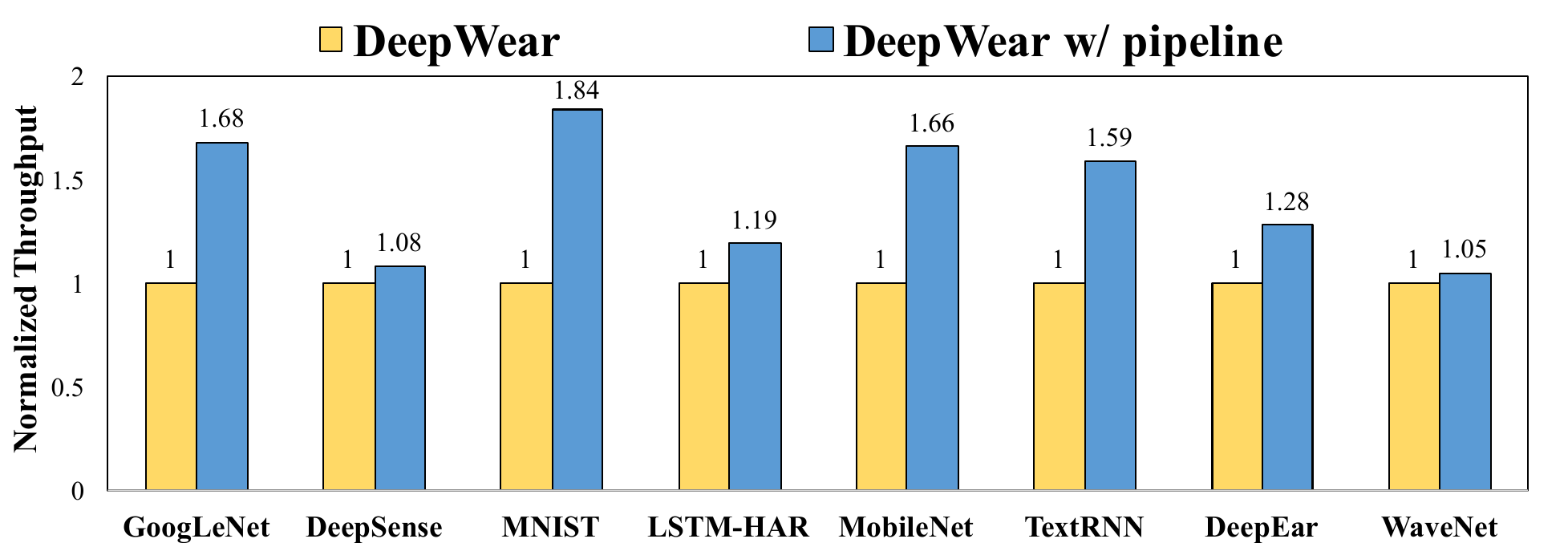}
	\caption{Throughput of \framework~with pipelined processing.
	\revise{Results are normalized by \framework~without pipelining. We use the Urbane LG and the Nexus 6 CPU-interactive to carry out this experiment.}
	}
	\label{fig:streaming}
\end{figure}

We also evaluate how the pipelining technique described in Section~\ref{sec:streaming} can help improve the throughput for streaming data.
As shown in Figure~\ref{fig:streaming}, applying pipelining in \framework~can help improve the overall throughput by 43.75\% averaged over the 8 models (the comparison baseline is the basic \framework~that treats each DL instance separately).
For some models such as \emph{MNIST}, the throughput improvement as high as 84\% can be achieved through pipelined processing.
We observe that the throughput boost depends on the processing time difference between
the wearable and the handheld. For models that exhibit large performance difference, applying pipelining achieves less improvement.
For example, running \emph{WaveNet} locally yields a latency of 7.7s on Urbane LG, almost 13 times higher than that achieved by offloading to Nexus CPU (0.54s).
As a result, applying pipelining increases the throughput by only 5\%.
This is because when the processing capabilities of the wearable and handheld differ significantly,
the little contribution of the weaker device (typically the wearable) makes pipelining fallback to the handheld-only strategy.
In contrast, for models that exhibit similar performance on the wearable and the handheld, pipelining leads to a much higher throughput improvement (84\% for \emph{MNIST} model).

\mysubsection{System Overhead}\label{sec:eval_overhead}
\framework incurs the computation overhead of executing the partition algorithm (Section~\ref{sec:decision}).
We have measured all 8 DL models under different configurations. The incurred overhead in terms of the fraction of latency is low, ranging from 0.49\% (\emph{GoogLeNet}) to 4.21\% (\emph{TextRNN}).
The reason for such low overhead is multifold. First, our heuristic-based algorithm, as presented in Section~\ref{sec:decision}, can reduce the computation complexity to almost $\mathcal{O}(n)$, where n is the number of DL model nodes. Second, the original DL computation is already heavy-load, making the overhead relatively trivial.

Another source of overhead comes from the \emph{System Profiler}. Our measurements indicate that such an overhead is non-trivial when the Bluetooth bandwidth is measured passively. \framework can optionally measure the Bluetooth bandwidth through active probing (Section~\ref{sec:implementation}). In that case the energy overhead is less than 5\% for the wearable. The overhead can be further reduced by using Bluetooth Low Energy (BLE) as instead of classic Bluetooth.

\mysection{Limitations}\label{sec:discuss}
We discuss some limitations of \framework and highlight several future research directions.

\noindent $\bullet$ \framework~currently focuses on the inference stage as opposed to the training stage.
In deep learning, which requires a pre-trained model integrated into applications or downloaded in advance.
Although performing inference may be sufficient for most applications, we also notice that in recent years there have emerged requirements to train (consume) the data immediately when it is produced on wearable devices.
\revise{We plan to extend \framework for the model training phase. The challenging issues for supporting the model training in \framework are in two folds. (1) Designing new latency and energy prediction models for the training procedure (e.g., based on the backpropagation algorithm). (2) Designing new offloading decision algorithms. Since the training phase requires both the forward and backward data flow in our model graph,
it is not immediately clear how much benefits partial offloading can reward, which shall be further explored in our future work.
}

\noindent $\bullet$ \framework makes partition decision based on two key metrics of user experience: the end-to-end latency and the energy consumption.
Besides them, other metrics such as memory usage (both average and peak) is another important metric that should be taken into account~\cite{conf/mobisys/HanSPAWK16}.
We plan to consider memory as a developer-specified policy similar to the latency (\textit{PropT}).
This extension can be integrated into \framework via a runtime predicator of memory usage for different partitions and a new set of APIs for developers.
%

\noindent $\bullet$ We have tested \framework~on only 3 devices (LG Urbane, Galaxy S2, and Nexus 6)
and 8 widely used DL models.
Though These models are representative and widely used, we plan to assess \framework~more broadly on other hardware platforms and DL models.

\revise{
\noindent $\bullet$
In many common scenarios, our offloading scheme in \framework exhibits unique advantages over the cloud-based offloading in terms of resource utilization, ubiquitous access, and privacy preservation. However, we should point out that
due to the limited processing capacity on handheld devices, \framework might still suffer from poor performance. For example, there are other concurrent workload (typically as background services) running on a handheld, or the DL task is too heavyweight to be carried out on a handheld.
Future work towards such problems includes adaptively offloading DL tasks to multiple personal mobile devices (e.g., a smartphone, a tablet, and a laptop), as well as using the cloud as an alternative offloading target when local resources are too insufficient while the privacy is not a critical concern.
}

\section{Conclusion}\label{sec:conclusion}
\revise{
Wearables provide an important data source for numerous applications that can be powered by DL tasks.
To enable efficient DL on wearables, We have developed \framework, a practical DL framework designed for wearables.
\framework can intelligently, transparently, and adaptively offload DL computations from a wearable to a paired handheld.
It introduces various novel techniques such as context-aware offloading, strategic model partition, and pipelining supports to better utilize the processing capacity from the wearable's nearby handhelds.
Our evaluation on COTS devices and popular DL models demonstrate \framework significantly outperforms both wearable-only and handheld-only approaches
by striking a better balance among the latency and the energy consumption on both sides.
%
We believe that the lessons learned from our \framework design and implementation will shed the light on developing future AI-powered systems on mobile, wearable, and Internet-of-things (IoT) applications.
%
%
In our future work, in addition to performing the tasks proposed in~\S\ref{sec:discuss}, we plan to apply \framework to develop real-world DL applications for off-the-shelf wearables. To help the research community reproduce our study, we will release the source code of \framework to be publicly available. 

} 

\bibliographystyle{IEEEtran}
\bibliography{section/ref}

\nocite{liu2007towards}


\end{document}